\documentclass[preprint,12pt,numbers,sort&compress]{elsarticle}

\usepackage{amssymb}
\usepackage{amsmath}
\usepackage{mathrsfs}
\usepackage{graphicx}
\usepackage{hyperref}
\hypersetup{hidelinks, colorlinks=true, allcolors=blue,}
\usepackage{soul, color, xcolor}
\soulregister\cite7
\soulregister\ref7

\journal{Elsevier}
\begin{document}

\begin{frontmatter}

\title{A fully one-sided diffuse-interface immersed boundary method for wall-modeled large-eddy simulation}

\author[label1]{Qian Mao\corref{cor1}}
\cortext[cor1]{Corresponding author: qian.mao@univ-amu.fr}
\author[label2]{Yoshiharu Tamaki}
\author[label1]{Song Zhao}
\author[label1]{Jérôme Jacob}
\author[label1]{Pierre Boivin}
\author[label1]{Julien Favier}

\affiliation[label1]{organization={Aix-Marseille Univ, CNRS, Centrale Med, M2P2}, city={Marseille}, country={France}}
\affiliation[label2]{organization={Department of Aeronautics and Astronautics, The University of Tokyo}, city={Tokyo 113-8656}, country={Japan}}

\begin{abstract}

Diffuse-interface immersed boundary methods (DIBMs) provide a simple and robust approach for simulating flows involving complex geometries. However, their inherent diffusion effect can contaminate the near-wall flow field and significantly degrade wall-shear-stress prediction in wall-modeled large-eddy simulation (WMLES). To address this limitation, we develop a WMLES approach based on a fully one-sided diffuse-interface immersed boundary method (FODIBM). By performing interpolation and spreading exclusively inside the immersed body, the proposed method removes the cross-boundary diffusion effect that adversely affects wall modeling in conventional DIBMs. A wall-shear-stress enforcement strategy is developed by coupling the wall-parallel immersed-boundary forcing with the wall shear stress predicted by an explicit wall model. In addition, a $\tau$-model based on the modeled turbulent shear-stress tensor is introduced to preserve the total shear-stress balance below the reference height. The method is first validated in high-Reynolds-number turbulent channel flows, showing good agreement with DNS data for the mean velocity, Reynolds shear stress, and skin-friction coefficient. Sensitivity studies with respect to grid resolution, reference height, wall inclination angle, and Reynolds number demonstrate the robustness of the method. Compared with the conventional DIBM, the proposed method substantially improves the overall prediction accuracy, particularly at low reference heights. The approach is further assessed for turbulent flow over a NACA23012 airfoil, where the predicted pressure distribution and lift coefficient agree well with experimental data.

\end{abstract}

\begin{keyword}

Diffuse-interface immersed boundary method, Lattice Boltzmann method, Wall-modeled large-eddy simulation

\end{keyword}

\end{frontmatter}

\section{Introduction}
\label{Intro}

Computational fluid dynamics (CFD) has become an indispensable tool for simulating and analyzing wall-bounded turbulent flows, which are ubiquitous in engineering applications. Among the available numerical approaches, large-eddy simulation (LES) resolves large-scale turbulent motions while modeling subgrid-scale effects, thereby offering a favorable compromise between accuracy and computational cost relative to direct numerical simulation (DNS) and Reynolds-averaged Navier-Stokes (RANS) simulations. Nevertheless, wall-resolved LES remains computationally expensive because of the stringent grid resolution required in the near-wall region. To mitigate this limitation, wall-modeled LES (WMLES) has been developed \cite{piomelli2008wall,bose2018wall}, in which inner-layer turbulence is modeled while the outer-layer turbulence is directly resolved. By relaxing the near-wall resolution requirements, WMLES substantially reduces computational cost while maintaining acceptable accuracy for high-Reynolds-number wall-bounded flows. Conventional WMLES implementations predominantly employ body-fitted meshes, which present significant challenges in generating high-quality grids suitable for eddy-resolving simulations, particularly for complex geometries. As an alternative, the immersed boundary method (IBM) allows simulations using simple Cartesian grids \cite{mittal2005immersed}, thereby greatly simplifying mesh generation and enhancing numerical flexibility.

Since the pioneering work of Peskin \cite{peskin1972flow}, a wide variety of immersed boundary methods (IBMs) have been developed. These methods are commonly classified into diffuse-interface (continuous forcing) and sharp-interface (discrete forcing) approaches, depending on how boundary conditions are imposed. In diffuse-interface IBMs, the forcing terms used to enforce boundary conditions are distributed from Lagrangian markers on the interface to the surrounding Eulerian grid via a regularized delta function \cite{peskin2002immersed,mao2023snap,mao2024hydrodynamic}. Representative formulations include penalty or feedback IBMs \cite{huang2011improved,mao2021hydrodynamic}, direct-forcing IBMs \cite{pinelli2010immersed,mao2026fluid}, and multi-direct forcing IBMs \cite{luo2007full,zhang2020relaxed}. The spreading of forcing terms inevitably introduces diffusion effects, which ensures a smooth flow field near the boundary but results in an inaccurate representation of the fluid-solid interface. In contrast, sharp-interface IBMs impose boundary conditions directly on the discrete Eulerian grid in the vicinity of the interface. Well-known examples include the ghost-cell method \cite{mittal2008versatile,jost2021direct}, the cut-cell method \cite{ingram2003developments,schneiders2013accurate}, and the immersed interface method \cite{li2003overview,brehm2013novel}. Compared with diffuse-interface IBMs, sharp-interface IBMs typically require more elaborate treatments near the interface, especially for complex three-dimensional geometries. These treatments may involve the classification of grid cells into fluid, solid, and boundary cells, or even cell trimming and reshaping. While such complex strategies can ensure high local accuracy, they may also introduce numerical discontinuities with spurious oscillations \cite{seo2011sharp,cai2021coupling}.

Due to the limited flexibility in refining Cartesian grids in the wall-normal direction near curved geometries, IBMs are generally not well suited for directly resolving the inner-layer turbulence. Instead, IBMs are more naturally coupled with WMLES (IBM-WMLES). Most recent advancements have focused on sharp-interface IBMs, as they generally provide higher accuracy \cite{tamaki2021wall,chen2014wall,roman2009simple,wang2024wall,van2022immersed,cai2023immersed,de2023coupling}. In such simulations, non-body-conforming grids exhibit conservation errors in the wall-parallel momentum \cite{capizzano2011turbulent,tamaki2017near,tamaki2021wall}, which compromise the accurate prediction of the shear-stress distribution and leads to deviations of the velocity profile from the logarithmic law. To address this issue, IBM-WMLES generally incorporates two aspects. The first aspect is the modeling of the unresolved near-wall turbulence by enforcing the shear-stress balance (i.e., the constant total shear-stress condition). A conventional approach is to introduce a modeled eddy viscosity based on the mixing-length model or the Spalart-Allmaras model \cite{chen2014wall,roman2009simple,wang2024wall,van2022immersed,cai2023immersed,de2023coupling}. However, this approach may induce a log-layer mismatch due to insufficient wall-normal turbulent mixing \cite{cai2023immersed,de2023coupling}. To overcome this limitation, Tamaki and Kawai \cite{tamaki2021wall} proposed introducing a modeled turbulent shear stress tensor instead of a modeled eddy viscosity, which effectively suppresses the log-layer mismatch. The second aspect involves applying wall boundary conditions obtained from the wall model, either as a wall shear stress or via velocity reconstruction. Direct imposition of the wall shear stress is generally not feasible because the immersed interface does not coincide with the Eulerian grid. An exception arises in cut-cell approaches based on the finite-volume method, in which the wall shear stress can be incorporated into the diffusive flux across the interface \cite{chen2014wall}. More commonly, the wall-parallel velocity at boundary points is reconstructed under the assumption that the near-wall velocity follows the wall-law profile \cite{roman2009simple,wang2024wall,van2022immersed,cai2023immersed,de2023coupling} or the simplified linear profile \cite{tamaki2021wall}. Several sharp-interface IBM-WMLES approaches have been applied to simulate airfoils and aircraft \cite{cai2023immersed,tamaki2024wall}. Despite these advances, sharp-interface IBMs generally require complex boundary treatments \cite{cai2021coupling}.

Efficient and robust diffuse-interface IBMs hold great potential for engineering applications of IBM-WMLES. However, challenges arise from the inherent diffusion effect of diffuse-interface IBMs. Two main issues can be identified. The first is that the IB forcing affects the accuracy of the wall model applied at reference or matching points near the wall. These points should be located sufficiently far from the wall while remaining within the valid region of the wall model, which generally requires higher grid resolution compared to sharp-interface IBMs. The second issue is that the IB forcing can disturb the shear-stress balance in the near-wall region, potentially causing failure of the WMLES \cite{wang2024wall}. Several attempts have been made to extend wall modeling techniques to diffuse-interface IBMs \cite{cheylan2021immersed,yan2024efficient,shi2019wall,ma2019dynamic,shi2022non,ma2021hybrid}. Most studies have concentrated on surface pressure or wake statistics, whereas detailed boundary-layer turbulence, including Reynolds stresses and skin friction, has received little attention. A limited number of studies have investigated Reynolds stresses in turbulent channel flows \cite{shi2022non,ma2021hybrid}, in which wall-aligned grids were employed. This grid arrangement minimizes conservation errors in the wall-parallel momentum and facilitates the application of wall-modeling strategies. Furthermore, diffuse-interface IBMs allow for coupling between the wall shear stress and the wall-parallel IB forcing. Shi et al. \cite{shi2019wall} proposed such an approach, while their method introduces an artificial parameter to control the spreading of the IB forcing.

To address these issues and enable the application of the diffuse-interface IBM to high-Reynolds-number WMLES, we couple a novel wall modeling approach with the fully one-sided diffuse-interface IBM (FODIBM) proposed by Mao et al. \cite{mao2026explicit}. This IBM has been successfully applied to the simulation of compressible flows involving detached and attached shocks as well as strong shock-obstacle interactions. The FODIBM performs interpolation and spreading exclusively on the interior side of the immersed boundary, thereby minimizing diffusion effect and making it particularly well suited for wall modeling. In the present study, an approach is proposed to enforce the wall-shear-stress condition by coupling the wall-parallel IB forcing with the wall shear stress predicted by an explicit wall model, without introducing any artificial parameters. To ensure shear-stress balance below the reference height, both a modeled eddy-viscosity approach and a modeled turbulent shear-stress tensor approach are considered. Several high-Reynolds-number validation cases are conducted to demonstrate the effectiveness of the proposed method for WMLES, including turbulent channel flows with both wall-aligned and wall-unaligned grid configurations, as well as turbulent flow over a NACA23012 airfoil at a finite angle of attack.

The remainder of the paper is structured as follows. Section \ref{Num} introduces the governing equations, the flow solver, and the proposed IBM-WMLES framework. Section \ref{res} presents and discusses the numerical validation. Finally, conclusions are summarized in Section \ref{conclu}.

\section{Numerical methods}
\label{Num}

In the present study, weakly compressible viscous flows are governed by the filtered compressible Navier-Stokes (N-S) equations in a conservative formulation. The N-S equations are solved using a hybrid lattice Boltzmann method (LBM) solver coupled with the total energy equation, as described in references \cite{farag2021unified,wissocq2022restoring}. To model the inner-layer turbulence, an explicit wall model that blends the linear and logarithmic laws is coupled with a fully one-sided diffuse-interface immersed boundary method. Two different approaches are introduced to balance the total shear stress in the near-wall region. The details are described below.

\subsection{Governing equations}
\label{sub_govequ}

The filtered N-S equations, which use Favre filtering ($\tilde{\phi}=\overline{\rho \phi}/\bar{\rho}$), are expressed as follows:

\begin{equation}
\frac{\partial\mathbf{Q}}{\partial t}+\frac{\partial\mathbf{F}_j}{\partial j}=\mathbf{S},
\end{equation}

\begin{equation}
\left.\mathbf{Q}=\left\{\begin{array}{c}\bar{\rho}\\\bar{\rho}\tilde{u}_i\\\bar{\rho} \tilde{E}\end{array}\right.\right\},\quad\mathbf{F}_j=\begin{Bmatrix}\bar{\rho}\tilde{u}_j\\\bar{\rho}\tilde{u}_{i}\tilde{u}_{j}+\bar{p}\delta_{ij}+\tau^{*}_{ij}\\(\bar{\rho} \tilde{E}+\bar{p})\bar{u}_j+\bar{u}_j\bar{\tau}_{ij}+q^{*}_j\end{Bmatrix},\quad\mathbf{S}=\begin{Bmatrix}0\\f_{ui}\\f_{E}\end{Bmatrix},
\end{equation}
where $\bar{\rho}$, $\tilde{u}_i$, $\bar{p}$, $f_{ui}$ and $f_E$ denote the filtered density, the filtered velocity, the filtered pressure, the momentum forcing term, and the energy forcing term, respectively. The total energy is defined as $\tilde{E}= \tilde{e} + \widetilde{u_iu_i}/2$, where $\tilde{e}=c_v \tilde{T}$ is the filtered internal energy and $\tilde{T}$ is the filtered temperature. Here, $c_v=R/(\gamma-1)$ and $c_p=\gamma R(\gamma-1)$ are the specific heat at constant volume and pressure, respectively, with $R$ being the specific gas constant and $\gamma=c_p/c_v$ the specific heat ratio. The N-S equations are closed by the ideal gas equation of state ($\bar{p}=\bar{\rho} R \tilde{T}$). The viscous stress tensor ($\tau^{*}_{ij}$) and the heat flux ($q^{*}_j$) are given by

\begin{equation}
\tau^{*}_{ij}=-(\bar{\tau}_{ij}+\tau_{ij}^{\mathrm{SGS}})=-2(\mu+\mu_t^{\mathrm{SGS}})[\tilde{S}_{ij}-\frac{1}{3}\tilde{S}_{kk}\delta_{ij}],
\label{eq_stresstensor1}
\end{equation}

\begin{equation}
q^{*}_{j}=-c_p\left( \frac{\mu}{\mathrm{Pr}}+\frac{\mu_t^{\mathrm{SGS}}}{\mathrm{Pr}_t}\right)\frac{\partial T}{\partial x_j},
\end{equation}
where $\mu=\bar{\rho}\nu$, $\delta_{ij}$, $\mathrm{Pr}$ and $\mathrm{Pr}_t$ are the molecular viscosity, the Kronecker delta function, the Prandtl number, and the turbulent Prandtl number, respectively. The filtered viscous stress tensor ($\bar{\tau}_{ij}$) is defined as

\begin{equation}
\bar{\tau}_{ij}=-2\mu\left(\tilde{S}_{ij}-\frac{1}{3}\tilde{S}_{kk}\delta_{ij}\right),
\end{equation}
where the resolved strain rate tensor ($\tilde{S}_{ij}$) is expressed as

\begin{equation}
\tilde{S}_{ij}=\frac{1}{2}\left(\frac{\partial\tilde{u}_{i}}{\partial x_{j}}+\frac{\partial\tilde{u}_{j}}{\partial x_{i}}\right).
\end{equation}

The subgrid-scale (SGS) stress ($\tau_{ij}^{\mathrm{SGS}}$) arising from the filtering process is given by

\begin{equation}
\tau_{ij}^{\mathrm{SGS}}=\bar{\rho}(\widetilde{u_{i}u_{j}}-\tilde{u}_{i}\tilde{u}_{j}).
\end{equation}

Employing the eddy viscosity closure for the subgrid stress tensor leads to

\begin{equation}
\tau_{ij}^{\mathrm{SGS}}=-2\mu_t^{\mathrm{SGS}}\left(\tilde{S}_{ij}-\frac{1}{3}\tilde{S}_{kk}\delta_{ij}\right)+\frac{1}{3}\tau_{kk}^{\mathrm{SGS}}\delta_{ij},
\end{equation}
where $\mu_t^{\mathrm{SGS}}=\bar{\rho}\nu_{t}^{\mathrm{SGS}}$ is the eddy viscosity. The isotropic contribution ($\frac{1}{3}\tau_{kk}^{\mathrm{SGS}}\delta_{ij}$) is neglected, as it becomes significant only in the vicinity of shocks, where numerical schemes introduce substantial artificial dissipation \cite{garnier2009large}. In this work, the Vreman subgrid model \cite{vreman2004eddy} is adopted to account for under-resolved eddies. This model retains the simplicity of the classical Smagorinsky model by relying solely on first-order velocity derivatives, while achieving performance comparable to that of dynamic models without requiring explicit filtering or ensemble averaging in homogeneous directions. The Vreman eddy viscosity on Cartesian grids is given by

\begin{equation}
\nu_{t}^{\mathrm{SGS}}=C\Delta^{2}|\tilde{S}^{*}|,\quad|\tilde{S}^{*}|=\sqrt{\frac{B_{\beta}}{A_{\beta}}},
\end{equation}
where $C \approx 2.5C_s^2$, with $C_s$ being the Smagorinsky constant. The filter size ($\Delta$) is taken as the local grid spacing. $A_{\beta}$ and $B_{\beta}$ denote the first and second invariants (traces) of the tensor $\boldsymbol{\beta}$, respectively:

\begin{equation}
\boldsymbol{\beta}=\boldsymbol{\alpha}^{\mathrm{T}}\cdot\boldsymbol{\alpha}=\alpha_{ki}\alpha_{kj},\quad\alpha_{ij}=\frac{\partial\tilde{u}_{j}}{\partial x_{i}},
\end{equation}

\begin{equation}
A_{\beta}=\mathrm{tr}(\boldsymbol{\beta})=\|\boldsymbol{\alpha}\|^{2}=\alpha_{ij}\alpha_{ij},
\end{equation}

\begin{equation}
B_{\beta}=\frac{1}{2}\left[(\mathrm{tr}(\boldsymbol{\beta}))^{2}-\mathrm{tr}(\boldsymbol{\beta}^{2})\right]=\beta_{11}\beta_{22}-\beta_{12}^{2}+\beta_{11}\beta_{33}-\beta_{13}^{2}+\beta_{22}\beta_{33}-\beta_{23}^{2}.
\end{equation}

\subsection{Hybrid lattice Boltzmann method}
\label{sub_LB}

In the LBM framework, the fluid is described by the particle distribution function $f(\boldsymbol{x},\boldsymbol{\zeta},t)$, which denotes the probability density of fluid particles with velocity $\boldsymbol{\zeta}$ at position $\boldsymbol{x}$ and time $t$. The evolution of $f(\boldsymbol{x},\boldsymbol{\zeta},t)$ is governed by the Boltzmann equation:

\begin{equation}
    \frac{\partial f}{\partial t} + \boldsymbol{\zeta} \cdot  \nabla f= \Gamma (f),
\label{eq_bol}
\end{equation}
where $\Gamma$ is the collision operator. The velocity space is discretized by a set of velocity vectors \{$\boldsymbol c_i$, $i=0$, \ldots, $Q-1$\}, where $Q$ denotes the total number of discrete velocities. The $D3Q19$ scheme is employed, and the corresponding discrete velocities are defined in Eq. (\ref{eq_cid3q19}) of \ref{app_d3q19}. The lattice Boltzmann equation discretized in velocity space, physical space and time is given by \cite{farag2021unified}

\begin{equation}
\begin{aligned}
    \bar{f}_i(\boldsymbol{x},t+\Delta t) &= f^{eq}_i(\boldsymbol{x}-\boldsymbol{c}_{i}\Delta t,t) + (1-\frac{\Delta t}{\bar{\tau}_{\scriptscriptstyle \mathrm{LB}}})\bar{f}^{neq}_i(\boldsymbol{x}-\boldsymbol{c}_{i}\Delta t,t) \\
    &+ \frac{\Delta t}{2}F_i(\boldsymbol{x}-\boldsymbol{c}_{i}\Delta t,t),
\end{aligned}
\label{eq_LB}
\end{equation} 
where $\bar{f}_i$ is defined as $\bar{f}_i=f_i -[(\Gamma_i+F_i)\Delta t]/2$. $\bar{\tau}_{\scriptscriptstyle \mathrm{LB}}$ is defined as $\bar{\tau}_{\scriptscriptstyle \mathrm{LB}}=\tau_{\scriptscriptstyle \mathrm{LB}} +\Delta t/2$, where $\tau_{\scriptscriptstyle \mathrm{LB}}=(\nu+\nu_{t}^{\mathrm{SGS}}) / c_s^2$ is the relaxation time and $\nu$ denotes the kinematic viscosity. $f^{eq}_i$ and $\bar{f}^{neq}_i$ denote the equilibrium and non-equilibrium distribution functions, as defined in Eqs. (\ref{eq_feq}) and (\ref{eq_fneq}) of Appendix \ref{app_d3q19}. The external body force term $F_i$ is defined as

\begin{equation}
    F_i = w_i\left[ \frac{\mathcal{H}_{i,\alpha}^{(1)}}{c_s^2}f_{u,\alpha} + \frac{\mathcal{H}^{(2)}_{i,\alpha\beta}}{2c_s^4}a_{\alpha\beta}^{F,(2)} \right],
\label{eq_fi}
\end{equation}
where $a_{\alpha\beta}^{F,(2)}$ is the Hermite moment defined in Eq. (\ref{eq_af2}) of \ref{app_d3q19}.

The macroscopic quantities $\rho$ and $\boldsymbol{u}$ are calculated by

\begin{equation}
    \rho = \sum_i \bar{f}_i,
\end{equation}

\begin{equation}
    \rho \boldsymbol{u}= \sum_i \boldsymbol{c}_{i}\bar{f}_i + \frac{\Delta t}{2}\boldsymbol{f}_{\boldsymbol{u}},
\end{equation}
where the momentum forcing term $\boldsymbol{f}_{\boldsymbol{u}}=\boldsymbol{f}^\mathrm{IB}$ is defined in Eq. (\ref{eq_fib}) of Section \ref{sub_IB}.

To facilitate the application of the hybrid LBM for compressible flows, Wissocq et al. \cite{wissocq2022restoring} proposed a scheme that couples the LBM with the total energy equation. The resulting discrete form of the energy equation, incorporating the forcing term, is expressed as \cite{wissocq2022restoring}:

\begin{equation}
\delta_t(\rho E)+\delta_\alpha F_{+\Delta\alpha/2}^{\rho E, NS}=f_{E},
\label{eq_energy}
\end{equation}
where $\delta_t$ and $\delta_\alpha$ denote discrete operators defined in Eq (\ref{eq_deltata}) of \ref{app_d3q19}. The flux $F_{+\Delta\alpha/2}^{\rho E, NS}$ is expressed as:
\begin{equation}
\begin{aligned}
F_{+\Delta\alpha/2}^{\rho E, NS} &= \mathscr{F}_{+\Delta\alpha/2}^*(\rho Hu_\alpha)+(h-k)\left[F_{+\Delta\alpha/2}^\rho-\mathscr{F}_{+\Delta\alpha/2}^*(\rho u_\alpha)\right] \\
&+u_\beta\left[F_{+\Delta\alpha/2}^{\rho u_\beta}-\mathscr{F}_{+\Delta\alpha/2}^*(\rho u_\alpha u_\beta+p\delta_{\alpha\beta})\right] \\
&-\lambda\delta_\alpha T(\boldsymbol{x}+\boldsymbol{e}_\alpha\Delta x,t),
\end{aligned}
\end{equation}
where $H=E+p/\rho$ is the total enthalpy. $h=\gamma RT/(\gamma-1)$ is the enthalpy and $k=u_\alpha^2/2$ is the kinetic energy. $\boldsymbol{e}_\alpha$ denotes the unity vector in the direction $\alpha$. $F_{+\Delta\alpha/2}^\rho$ and $F_{+\Delta\alpha/2}^{\rho u}$ are mass and momentum fluxes defined in Eqs. (\ref{eq_fluxx}), (\ref{eq_fluxy}) and (\ref{eq_fluxz}) of \ref{app_d3q19}. $\mathscr{F}_{+\Delta\alpha/2}^*$ is the linear function defined in Eq. (\ref{eq_linear}) of \ref{app_d3q19}, determined using the MUSCL-Hancock scheme \cite{wissocq2022restoring}. The energy forcing term $f_{E}$ including the kinetic energy part $w^\mathrm{IB}$ and the internal energy part $q^\mathrm{IB}$ is defined in Eqs. (\ref{eq_wib}) and (\ref{eq_qib}) of Section \ref{sub_IB}. After solving the Eq. (\ref{eq_energy}), the temperature is obtained from $E=||\boldsymbol{u}||^2/2+C_v T$ and the pressure can be calculated by $p=\rho R T$.

\subsection{Diffuse-interface immersed boundary method}
\label{sub_IB}

\begin{figure}
\centerline{\includegraphics[width=0.9\linewidth]{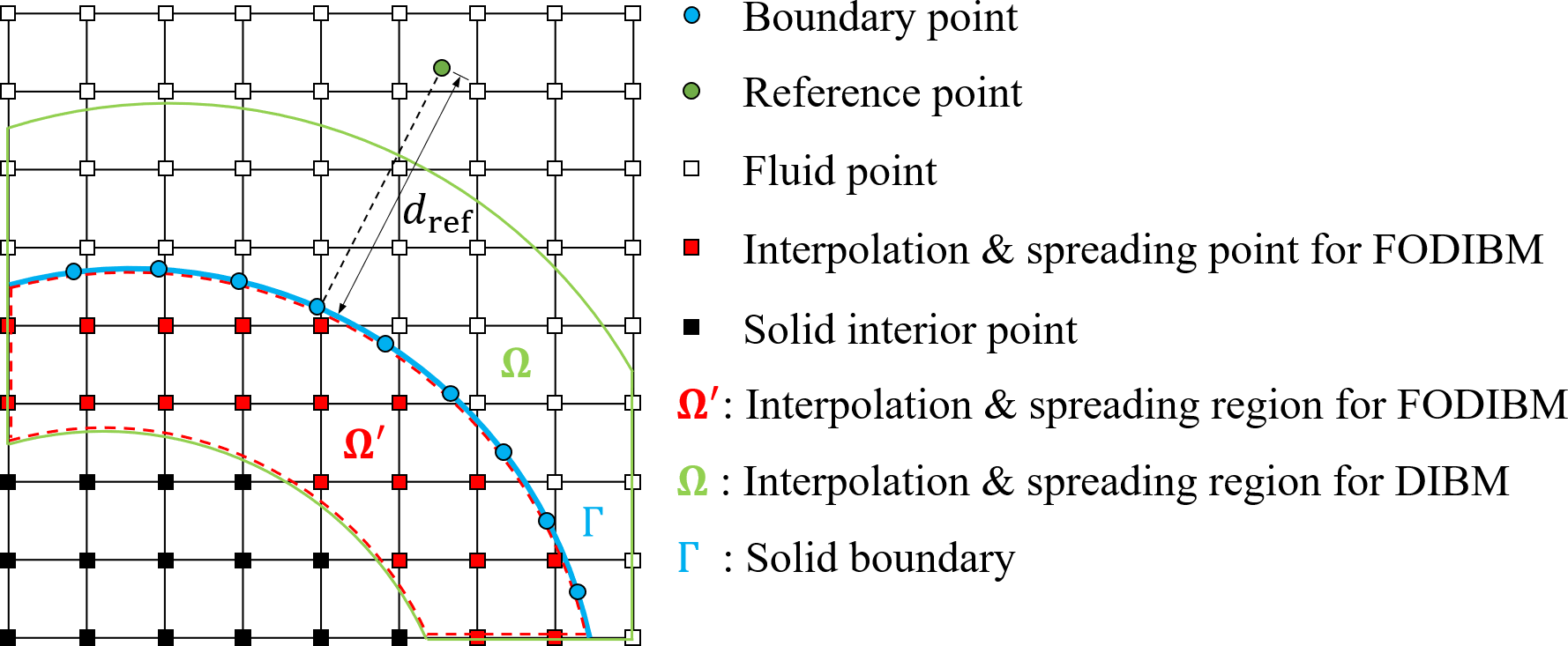}}
\caption{Schematic of the DIBM and FODIBM. Each boundary point is associated with a reference point at a distance $d_\mathrm{ref}$ along the interface normal for the wall model.}
\label{fig_one-side}
\end{figure}

Here, both the conventional diffuse-interface immersed boundary method (DIBM) and the fully one-sided diffuse-interface immersed boundary method (FODIBM) proposed by Mao et al. \cite{mao2026explicit} are introduced. Figure \ref{fig_one-side} schematically illustrates the DIBM and the FODIBM. The solid boundary ($\Gamma$) is discretized by a set of Lagrangian points ($\boldsymbol{X}_l$) defined on a curvilinear coordinate system $(q,r,s)$, while the fluid domain is discretized using a fixed Cartesian mesh. The set $\Omega$ denotes all Eulerian points ($\boldsymbol{x}_i$) located both inside and outside the solid boundary, whereas $\Omega^{\prime}$ denotes the subset of Eulerian points located strictly inside the solid boundary. In the DIBM, both the interpolation and spreading operations are performed over the entire set $\Omega$, i.e., on grid points located on both sides of the solid boundary. This treatment introduces a diffusion effect across the boundary, which reduces the accuracy of the flow-field prediction in the near-boundary region and prevents the correct imposition of Neumann boundary conditions. In contrast, in the FODIBM, both interpolation and spreading are strictly confined to $\Omega^{\prime}$. This one-sided strategy effectively eliminates the diffusion-related issues inherent to the DIBM. As a result, the FODIBM exhibits improved performance, particularly in simulations of compressible flows \cite{mao2026explicit}. In the present study, the FODIBM is also shown to perform better in high-Reynolds-number WMLES, as demonstrated in Section \ref{channel}. The reference point is defined for the wall model introduced in Section \ref{sub_WM}. The detailed formulations of both the DIBM and the FODIBM are presented below.

The IB forcing is computed based on the predicted flow variables interpolated at the Lagrangian points. Subsequently, the forcing term to be incorporated into the governing equations is obtained by spreading the IB force onto the surrounding Eulerian points. We first consider the DIBM, in which the interpolation step is formulated as

\begin{equation}
    \boldsymbol{M}^*_l=\mathcal{I}[\boldsymbol{m}]_l=\sum_{\Omega} \boldsymbol{m}_i\delta(\boldsymbol{x}_{i}-\boldsymbol{X}_{l})\Delta x^3,
\label{eq_int1}
\end{equation}
where $\boldsymbol{x}_i \in \Omega$ indicates the fluid points both inside and outside the solid boundary. $\boldsymbol{m}$ denotes flow variables at Eulerian points, such as velocity, density and temperature. $\boldsymbol{M}^*$ is the predicted flow variables at Lagrangian points. $\mathcal{I}[\bullet]_l$ is the interpolation operator. The spreading step is expressed as

\begin{equation}
    \boldsymbol{g}_i^{\mathrm{IB}}=\mathcal{S}[\boldsymbol{G}^{\mathrm{IB}}]_i=\sum_{\Gamma} \boldsymbol{G}_l^{\mathrm{IB}}\delta(\boldsymbol{x}_{i}-\boldsymbol{X}_{l})W_l,
\label{eq_spr1}
\end{equation}
where $\boldsymbol{X}_l \in \Gamma$. $\boldsymbol{g}^{\mathrm{IB}}$ is the IB forcing term which will be incorporated into the governing equations, such as the momentum forcing term ($\boldsymbol{f}^{\mathrm{IB}}$) and the energy forcing term (includes kinetic energy part $w^{\mathrm{IB}}$ and internal energy part $q^{\mathrm{IB}}$). $\boldsymbol{G}^{\mathrm{IB}}$ (includes $\boldsymbol{F}^{\mathrm{IB}}$, $W^{\mathrm{IB}}$ and ${Q}^{\mathrm{IB}}$) is the IB force at Lagrangian points. $W_l=\Delta q \Delta r \Delta s=\Delta S \Delta x$ is the Lagrangian weight, where $\Delta S$ is the area of a Lagrangian grid. An improved Lagrangian weight recovering the interpolation/spreading reciprocity condition will be defined hereafter, which decreases errors in both Dirichlet and Neumann boundary conditions \cite{mao2026explicit}.

For the FODIBM, both the interpolation and spreading steps are corrected by introducing a scaling factor ($\phi_l$) to recover the equivalence between the Eulerian and Lagrangian variables \cite{mao2026explicit}, since the interpolation domain is reduced from $\Omega$ to $\Omega^{\prime}$:

\begin{equation}
    \boldsymbol{M}^*_l=\mathcal{I}[\boldsymbol{m}]_l=\sum_{\Omega ^ {\prime}} \phi_l\boldsymbol{m}_i\delta(\boldsymbol{x}_{i}-\boldsymbol{X}_{l})\Delta x^3,
\label{eq_int2}
\end{equation}

\begin{equation}
    \boldsymbol{g}_i^{\mathrm{IB}}=\mathcal{S}[\boldsymbol{G}^{\mathrm{IB}}]_i=\sum_{\Gamma} \phi_l\boldsymbol{G}_l^{\mathrm{IB}}\delta(\boldsymbol{x}_{i}-\boldsymbol{X}_{l})W_l,
\label{eq_spr2}
\end{equation}
where $\phi_l$ is defined as

\begin{equation}
\phi_l=\frac{1}{\sum_{\Omega ^ {\prime}}\delta(\boldsymbol{x}_{i}-\boldsymbol{X}_{l})\Delta x^3}.
\end{equation}

The derivation of $\phi_l$ is presented in our previous work \cite{mao2026explicit}. For the conventional direct-forcing IBM, the interpolation/spreading is non-reciprocal ($\mathcal{I}[\mathcal{S}[\boldsymbol{G}^{\mathrm{IB}}]]\neq\boldsymbol{G}^{\mathrm{IB}}$) in explicit implementations, which causes the boundary error as demonstrated by Gsell and Favier \cite{gsell2021direct}. This error corresponds to discrepancies between the target and the realized values at the boundary. To enforce reciprocity between interpolation and spreading, an improved Lagrangian weight is introduced, such that $\mathcal{I}[\mathcal{S}[\boldsymbol{G}^{\mathrm{IB}}]]\approx \boldsymbol{G}^{\mathrm{IB}}$. In other words, spreading $\boldsymbol{G}^{\mathrm{IB}}$ onto the Eulerian field and then interpolating it back yields approximately the same value. The improved Lagrangian weight ($W_l$) normalized by $\Delta x^3$ is defined as

\begin{equation}
    W_l = \frac{1}{\sum_i\sum_k[\phi_k\delta(\boldsymbol{x}_i-\boldsymbol{X}_k)][\phi_l\delta(\boldsymbol{x}_i-\boldsymbol{X}_l)]}.
\label{eq_wpod}
\end{equation}

The derivation of $W_l$ is presented in our previous work \cite{mao2026explicit}. The value of $\phi_l$ is set to unity for the DIBM. The Lagrangian weight is determined exclusively by the geometry of the boundary and its discretization. As it depends only on a limited number of neighboring Eulerian and Lagrangian points, the associated computational cost is minimal. For stationary bodies, this weight remains constant and can be computed once during the initialization stage, making its contribution to the overall computational cost negligible.

The delta function ($\delta$) for interpolation and spreading is expressed as

\begin{equation}
\delta(\boldsymbol{x}_{i}-\boldsymbol{X}_{l})=\frac{1}{\Delta x^3}\widetilde{\delta}\left(\frac{{x}_{i}-X_{l}}{\Delta x}\right)\widetilde{\delta}\left(\frac{{y}_{i}-Y_{l}}{\Delta x}\right)\widetilde{\delta}\left(\frac{{z}_{i}-Z_{l}}{\Delta x}\right),
\end{equation}

\begin{equation}
\widetilde{\delta}(r)=\begin{cases}\frac{1}{2d}[1+\cos(\frac{\pi r}{d})] &|r|\leq d,\\0 &|r|>d,\end{cases}
\end{equation}
where $d=2$ is the radius of the delta function.

In the present study, $\eta$ and $\xi$ are the wall-normal and -parallel coordinates, respectively. The IB forcing term $\boldsymbol{g}^{\mathrm{IB}}$ comprises the momentum forcing term ($\boldsymbol{f}^{\mathrm{IB}}$) and the energy forcing terms (${w}^{\mathrm{IB}}$ and ${q}^{\mathrm{IB}}$). A non-penetration condition is enforced in the wall-normal direction, resulting in the wall-normal momentum IB force \cite{cheylan2023analysis,mao2026explicit}:

\begin{equation}
\boldsymbol{F}^{\mathrm{IB}}_{\eta l}=\frac{2[\rho^*\boldsymbol{u}^t_{\eta}- (\rho\boldsymbol{u}_{\eta})^*]}{\Delta t},
\label{eq_fib}
\end{equation}
where $\boldsymbol{u}^t_{\eta}=0$ is the target wall-normal velocity at the Lagrangian point. $\rho^*$ and $(\rho\boldsymbol{u}_{\eta})^*$ are predicted density and wall-normal momentum obtained by Eq. (\ref{eq_int2}). $\boldsymbol{f}^{\mathrm{IB}}_{\eta}$ is then obtained by spreading $\boldsymbol{F}^{\mathrm{IB}}_{\eta}$ to the surrounding Eulerian points by Eq. (\ref{eq_spr2}).

In the present study, the wall-parallel momentum IB force is coupled with the wall model, which is introduced in Section \ref{sub_WM2}. For comparison, a case with a no-slip boundary condition is also considered, yielding the following wall-parallel momentum IB force with $\boldsymbol{u}^t_{\xi}=0$ (target wall-parallel velocity):

\begin{equation}
\boldsymbol{F}^{\mathrm{IB}}_{\xi l}=\frac{2[\rho^*\boldsymbol{u}^t_{\xi}- (\rho\boldsymbol{u}_{\xi})^*]}{\Delta t},
\label{eq_fib_xi}
\end{equation}

The energy forcing term at the Lagrangian point can be separated into kinetic and thermal forcings \cite{lerogeron2025numerical}:
\begin{equation}
    {W}^{\mathrm{IB}}_l=\frac{\rho^*[(\boldsymbol{u}^t)^2- (\boldsymbol{u}^*)^2]}{2\Delta t},
\label{eq_wib}
\end{equation}

\begin{equation}
    {Q}^{\mathrm{IB}}_l=\frac{C_v\rho^*(T^t- T^*)}{\Delta t},
\label{eq_qib}
\end{equation}
where $T^t$ is the target temperature at the Lagrangian point. $\boldsymbol{u}^*$ and $T^*$ are predicted velocity and temperature obtained by Eq. (\ref{eq_int2}). ${w}^{\mathrm{IB}}$ and ${q}^{\mathrm{IB}}$ is then obtained by spreading ${W}^{\mathrm{IB}}$ and ${Q}^{\mathrm{IB}}$ to the surrounding Eulerian points by Eq. (\ref{eq_spr2}). The values of $T^t$ satisfying either Dirichlet or Neumann boundary conditions can be specified following the approach presented in our previous work \cite{mao2026explicit}. In the present study, an isothermal boundary condition ($T^t = T_0$, where $T_0$ denotes the initial temperature) is imposed, and temperature variations are negligible at low Mach numbers ($Ma < 0.2$). As a result, the density and viscosity fields remain nearly uniform.

\subsection{Wall modelling with immersed boundary method}
\label{sub_WM}

\subsubsection{Blended explicit wall model}

When the near-wall grid resolution is insufficient to support LES, a wall model can be employed to provide an appropriate wall shear stress ($\tau_{w}$) as a boundary condition. To predict $\tau_{w}$, a reference point is defined for each Lagrangian point at a distance $d_{\mathrm{ref}}$ along the interface normal as illustrated in Fig. \ref{fig_one-side}. In the present study, the explicit wall model proposed by Cai et al. \cite{cai2023immersed} is adopted. This model blends the linear law and the logarithmic law to obtain a unified velocity profile. The maximum error of the model in the blending region is below $0.3\%$ and decreases to $0.01\%$ in the logarithmic layer, thereby achieving both high accuracy and low computational cost. The local Reynolds number is defined as

\begin{equation}
Re_{y}=\frac{u_{\xi}y}{\nu}=u_{\xi}^{+}y^{+},
\end{equation}
where $y=d_{\mathrm{ref}}$, $u_{\xi}^{+}=u_{\xi}/u_{\tau}$, and $y^{+}=d_{\mathrm{ref}}u_{\tau}/\nu$, with $u_{\tau}$ being the friction velocity. $u_{\xi}$ is the wall-parallel velocity on the reference point, which can be obtained by interpolation using Eq. (\ref{eq_int1}). $y^{+}$ is calculated by

\begin{equation}
y^{+}=\left[1-\tanh(\frac{Re_{y}}{s})\right]^{p}(Re_{y})^{1/2}+\left[\tanh(\frac{Re_{y}}{s})\right]^{p}\frac{1}{E}e^{W(\kappa ERe_{y})},
\label{eq_yplus}
\end{equation}
where $s=180.8$, $p=0.789$, $E=11.27$, $\kappa=0.41$. The sixth-order truncated Lambert $W$ function is defined as

\begin{equation}
W(x)\approx\ln(\frac{x}{\ln(\frac{x}{\ln(x)})})=\ln(x)-\ln(\ln(x)-\ln(\ln(x))).
\end{equation}

The friction velocity and wall shear stress can be obtained by

\begin{equation}
u_\tau=y^+\nu/d_{\mathrm{ref}},
\label{eq_utau}
\end{equation}

\begin{equation}
\tau_{w} = \rho u_\tau^2.
\label{eq_tauw}
\end{equation}

\subsubsection{Coupling between the wall-parallel IB force and the wall model}
\label{sub_WM2}

\begin{figure}
\centerline{\includegraphics[width=0.45\linewidth]{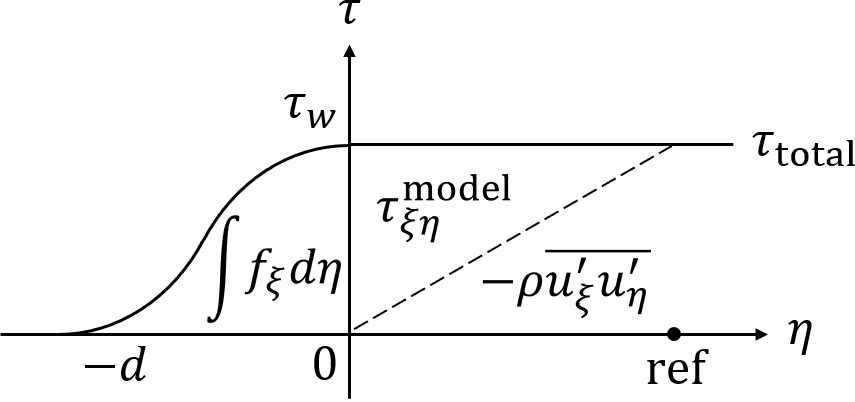}}
\caption{Schematic of the shear-stress balance.}
\label{fig_WM}
\end{figure}

In the present study, to enforce a wall-shear-stress condition at the wall, an approach is proposed to couple the wall-parallel IB force ($\boldsymbol{F}^{\mathrm{IB}}_{\xi l}$) with the wall shear stress ($\tau_w$) predicted by the wall model. For the FODIBM, $\boldsymbol{F}^{\mathrm{IB}}_{\xi l}$ is distributed over the region $\eta \in (-d, 0)$, and the integration of its Eulerian counterpart $f^{\mathrm{IB}}_{\xi}$ in the wall-normal direction is expressed as

\begin{equation}
    \int_{-d}^0 f^{\mathrm{IB}}_{\xi} d\eta = \tau_{w},
\label{eq_intfib}
\end{equation}
which enforces the wall-shear-stress condition at the wall, as illustrated in Fig. \ref{fig_WM}. The discretized form of Eq. (\ref{eq_intfib}) is given by ($\Delta \eta=\Delta \xi=\Delta x$)

\begin{equation}
    \sum_{\eta} f^{\mathrm{IB}}_{\xi} \Delta x = \tau_{w}.
\label{eq_sumfib1}
\end{equation}

The summation of $\boldsymbol{f}^{\mathrm{IB}}_{\xi}$ in the wall-parallel and wall-normal directions is then given by

\begin{equation}
    \sum_{\xi} \sum_{\eta} \boldsymbol{f}^{\mathrm{IB}}_{\xi} = -\frac{\sum_{\xi}\tau_{w i}}{\Delta x} \boldsymbol{\xi},
\label{eq_sumfib2}
\end{equation}
where $\boldsymbol{\xi}$ denotes the wall-parallel unit vector. In the present study, it is computed as $\boldsymbol{u}_{\xi} / |\boldsymbol{u}_{\xi}|$. According to the equivalence between the Eulerian and Lagrangian descriptions, the summation of the wall-parallel momentum IB force is expressed as

\begin{equation}
    \sum \boldsymbol{F}^{\mathrm{IB}}_{\xi l} \Delta S \Delta x = \sum_{\xi} \sum_{\eta} \boldsymbol{f}^{\mathrm{IB}}_{\xi} \Delta x ^3= -\sum_{\xi}\tau_{w i}\Delta x ^2 \boldsymbol{\xi},
\label{eq_fibtang1}
\end{equation}
where $\Delta S$ denotes the area of a Lagrangian grid cell, which may differ from $\Delta x^2$. The quantity $\tau_{w i}$ represents the wall shear stress per unit area ($\Delta x^2$) in the Eulerian space. The equivalent form of $\sum_{\xi} \tau_{w i} \Delta x^2 \boldsymbol{\xi}$ in the Lagrangian space can be written as $\sum \tau_{w l} \Delta S \boldsymbol{\xi}$. Therefore, Eq. (\ref{eq_fibtang1}) can be reformulated as

\begin{equation}
    \sum \boldsymbol{F}^{\mathrm{IB}}_{\xi l} = -\frac{\sum\tau_{w l} \boldsymbol{\xi}}{\Delta x},
\label{eq_fibtang2}
\end{equation}

To satisfy Eq. (\ref{eq_fibtang2}) while reducing implementation complexity, the instantaneous local value of $\tau_w$ is used in the calculation of $\boldsymbol{F}^{\mathrm{IB}}_{\xi l}$:

\begin{equation}
    \boldsymbol{F}^{\mathrm{IB}}_{\xi l} =-\frac{\tau_{wl}}{\Delta x} \boldsymbol{\xi},
\label{eq_fibtang4}
\end{equation}
where $\tau_{w l}$ is the wall shear stress at the Lagrangian point $l$, predicted by the wall model at the corresponding reference point. The reciprocity condition is no longer required for the wall-parallel momentum forcing term, since $\boldsymbol{F}^{\mathrm{IB}}_{\xi l}$ is directly linked to the wall shear stress and the concept of a velocity boundary-condition error does not apply. Consequently, the Lagrangian weight $W_l$ reduces to $\Delta S \Delta x$ for $\boldsymbol{F}^{\mathrm{IB}}_{\xi l}$.

\subsubsection{Shear-stress balance in the near-wall region}

In addition to enforcing the wall-shear-stress condition at the wall via the IB force, ensuring shear-stress balance in the near-wall region is also important. By integrating the wall-parallel momentum-conservation equation in the wall-normal direction, we obtain ($\Delta>0$):

\begin{equation}
\int_{0}^{\Delta} \frac{d \bar{\tau}}{d \eta}d\eta=
\int_{0}^{\Delta} \rho\left[\bar{u}_\xi \frac{\partial \bar{u}_\xi}{\partial \xi} 
+ \bar{u}_\eta \frac{\partial \bar{u}_\xi}{\partial \eta}\right]d\eta
+\int_{0}^{\Delta} \frac{\partial \bar{p}}{\partial \xi}d\eta,
\label{eq_integmoment1}
\end{equation}
where the overline denotes ensemble averaging, and $\bar{\tau}$ is the total shear stress, comprising both viscous and Reynolds shear stress components. $f_\xi^{\mathrm{IB}}$ is omitted as it is applied only inside the boundary ($\eta < 0$). The inertial terms and the pressure gradient are negligible in the near-wall region of the boundary layer. Consequently, Eq. (\ref{eq_integmoment1}) can be simplified as

\begin{equation}
\int_{0}^{\Delta} \frac{d \bar{\tau}}{d \eta}d\eta = \bar{\tau}(\Delta)-\bar{\tau}({0}) \approx 0.
\label{eq_integmoment2}
\end{equation}

$f_\xi^{\mathrm{IB}}$ enforces the wall shear stress at the boundary ($\eta=0$), i.e. $\bar{\tau}({0})=\bar{\tau}_w$. Substituting $\bar{\tau}(\Delta)=-\rho\overline{u_{\xi}^{\prime}u_{\eta}^{\prime}}+(\mu+\bar{\mu}_{t}^{\mathrm{SGS}})d\bar{u}_{\xi}/d\eta$ into Eq. (\ref{eq_integmoment2}) yields:

\begin{equation}
-\rho\overline{u_{\xi}^{\prime}u_{\eta}^{\prime}}+(\mu+\bar{\mu}_{t}^{\mathrm{SGS}})\frac{d\bar{u}_{\xi}}{d\eta} \approx \bar{\tau}_w,
\end{equation}
which means that the total shear stress is approximately equal to the mean wall shear stress, i.e. the near-wall shear-stress balance. When the near-wall grid resolution is insufficient, the Reynolds shear stress ($-\rho \overline{u_{\xi}^{\prime} u_{\eta}^{\prime}}$) is typically underestimated. Consequently, the velocity gradient in the near-wall region increases to compensate for the missing shear stress, leading to deviations of the velocity profile from the classical logarithmic law.

To properly maintain the total shear-stress balance, the modeled turbulent shear stress $\tau_{\xi \eta}^{\mathrm{model}}$ is introduced in the near-wall region. With the inclusion of $\tau_{\xi \eta}^{\mathrm{model}}$, the total shear-stress balance can be expressed as

\begin{equation}
-\rho\overline{u_{\xi}^{\prime}u_{\eta}^{\prime}}+(\mu+\bar{\mu}_{t}^{\mathrm{SGS}})\frac{d\bar{u}_{\xi}}{d\eta}+\tau_{\xi \eta}^{\mathrm{model}}\approx\bar{\tau}_w,
\end{equation}
where the viscous and SGS shear stresses ($(\mu+\bar{\mu}_{t}^{\mathrm{SGS}}){d\bar{u}_{\xi}}/{d\eta}$) is negligible throughout the boundary layer. This results from the intentionally reduced near-wall velocity gradient induced by $\tau_{\xi \eta}^{\mathrm{model}}$ and the wall-model-linked wall-parallel IB force, in contrast to the conventional no-slip boundary condition. At the wall ($\eta = 0$), the resolved Reynolds shear stress ($-\rho \overline{u_{\xi}^{\prime} u_{\eta}^{\prime}}$) is approximately zero due to the non-penetration condition ($u_{\eta} = 0$). Therefore, $\tau_{\xi \eta}^{\mathrm{model}}$ should equal to $\bar{\tau}_w$ at $\eta = 0$. Beyond the reference height ($\eta = d_{\mathrm{ref}}$), the energy-carrying turbulence is fully resolved by LES, and $\tau_{\xi \eta}^{\mathrm{model}}$ decreases to zero at $\eta = d_{\mathrm{ref}}$. Here, the profile of $\tau_{\xi \eta}^{\mathrm{model}}$ between the wall and the reference height is assumed to be linear, following the study of Tamaki and Kawai \cite{tamaki2021wall}. A schematic illustrating the shear-stress balance is shown in Fig. \ref{fig_WM}.

The viscous stress tensor in Eq. (\ref{eq_stresstensor1}) with modeled turbulent shear stress is rewritten as

\begin{equation}
\tau^{*}_{ij}=-2(\mu+\mu_t^{\mathrm{SGS}})[\tilde{S}_{ij}-\frac{1}{3}\tilde{S}_{kk}\delta_{ij}] +\tau_{ij}^{\mathrm{model}}.
\label{eq_stresstensor2}
\end{equation}

To determine the value of $\tau_{ij}^{\mathrm{model}}$, two approaches are introduced. The first approach increases the eddy viscosity isotropically in the near-wall region and is referred to as the $\mu$-model in the present study. This approach represents a type of hybrid RANS-LES method, and $\tau_{ij}^{\mathrm{model}}$ is calculated by:

\begin{equation}
\tau_{ij}^{\mathrm{model}}=-2\mu_t^{\mathrm{RANS}}(\tilde{S}_{ij}-\frac{1}{3}\tilde{S}_{kk}\delta_{ij})f_s,
\label{eq_taumu}
\end{equation}
where $f_s$ is a blending function that takes the value of unity at the wall and zero at the reference height. In the present study, despite the possibility of alternative choices, this simple linear form is adopted for convenience:

\begin{equation}
f_s=\max\left(\frac{d_{\mathrm{ref}}-\eta}{d_{\mathrm{ref}}},0\right),
\end{equation}
where $\eta$ denotes the wall-normal distance from the wall. To incorporate the $\mu$-model into the hybrid LBM framework, $\nu_t^{\mathrm{RANS}}=\mu_t^{\mathrm{RANS}}/\rho$ is added to the relaxation time ($\tau_{\scriptscriptstyle \mathrm{LB}}$), i.e. $\tau_{\scriptscriptstyle \mathrm{LB}}=(\nu+\nu_{t}^{\mathrm{SGS}}+\nu_t^{\mathrm{RANS}}) / c_s^2$. The calculation of $\mu_t^{\mathrm{RANS}}$ will be introduced later.

The $\mu$-model is a conventional approach for maintaining the total shear-stress balance in WMLES. However, it can induce a log-layer mismatch due to insufficient wall-normal turbulent mixing in the near-wall region, resulting from the excessive eddy viscosity \cite{davidson2006hybrid,piomelli2008wall}. This issue commonly arises in hybrid RANS-LES simulations \cite{davidson2003hybrid,hamba2003hybrid,cai2023immersed}. To address this problem, the present study adopts the method proposed by Tamaki and Kawai \cite{tamaki2021wall}, referred to here as the $\tau$-model:

\begin{equation}
\tau_{ij}^{\mathrm{model}}=\tau_w(\xi_i \eta_j+\eta_i \xi_j)f_s,
\label{eq_tautau1}
\end{equation}
where $\xi_i$ and $\eta_i$ denote the wall-parallel and wall-normal unit vectors, respectively. This method adds only the shear-stress component to the viscous stress tensor while keeping its diagonal components unchanged, thereby rendering the tensor anisotropic. Such a strategy minimizes adverse effects on wall-normal turbulence mixing.

Equation (\ref{eq_tautau1}) requires the wall shear stress $\tau_w$ at the closest wall location, which complicates its implementation. To avoid searching for the nearest wall from every fluid point, an alternative estimation method for $\tau_w$ is employed. This method is based on the observation that the total shear stress remains approximately constant within the inner layer of the turbulent boundary layer. In a statistically averaged flow field, this assumption can be expressed as

\begin{equation}
\tau_{\mathrm{total}}\approx(\mu+\mu_{t}^{\mathrm{RANS}})\frac{d\bar{u}_{\xi}}{d\eta}\approx\tau_w,
\end{equation}
where the RANS eddy viscosity ($\mu_t^{\mathrm{RANS}}$) is approximated using the mixing-length theory \cite{baldwin1978thin}:

\begin{equation}
\mu_{t}^{\mathrm{RANS}}\approx\rho(\kappa \eta)^2\frac{d\bar{u}_{\xi}}{d\eta}.
\label{eq_murans}
\end{equation}

At the reference height, it follows that

\begin{equation}
\tau_{\mathrm{total}}\approx\rho(\kappa d_{\mathrm{ref}})^2\left(\frac{d\bar{u}_{\xi}}{d\eta}\bigg|_{\mathrm{ref}}\right)^2\approx\tau_w,
\label{eq_tauw_ref}
\end{equation}
where the viscous shear stress term is neglected because the reference point lies within the nearly inviscid logarithmic layer. Under the assumption of a linear velocity profile below the reference height, the velocity gradient at the reference point is approximated by the local gradient. Combining Eqs. (\ref{eq_tautau1}) and (\ref{eq_tauw_ref}) gives

\begin{equation}
\tau_{ij}^{\mathrm{model}}=\rho(\kappa d_{\mathrm{ref}})^2\frac{d\bar{u}_{\xi}}{d\eta}\left|\frac{d\bar{u}_{\xi}}{d\eta}\right|(\xi_i \eta_j+\eta_i \xi_j)f_s.
\label{eq_tautau2}
\end{equation}

To incorporate the $\tau$-model into the hybrid LBM framework, a term of $a_{\alpha\beta}^{\mathrm{model},(2)}=-\tau_{\alpha\beta}^{\mathrm{model}}/\tau_{\scriptscriptstyle \mathrm{LB}}$ is added to the Hermite moment ($a_{\alpha\beta}^{F,(2)}$) in Eq. (\ref{eq_fi}). The modified Hermite moment is then expressed as
\begin{equation}
a_{\alpha\beta}^{F,(2)*}=a_{\alpha\beta}^{F,(2)}+a_{\alpha\beta}^{\mathrm{model},(2)}.
\end{equation}

\subsection{Computational procedure}
\label{sub_Algorithm}

The computational procedure of the present FODIBM for WMLES at each time step is summarized as follows:

1. Compute the predicted variables $\rho^*$, $\boldsymbol{u}^*$, and $T^*$ using Eq. (\ref{eq_int2}). The flow variables from the previous time step are treated as intermediate variables.

2. Evaluate the velocity $\boldsymbol{u}$ at the reference point using Eq. (\ref{eq_int1}).

3. Predict the wall shear stress $\tau_w$ using the wall model through Eqs. (\ref{eq_yplus}), (\ref{eq_utau}), and (\ref{eq_tauw}).

4. Compute the modeled shear stress $\tau_{ij}^{\mathrm{model}}$ using either Eq. (\ref{eq_taumu}) or Eq. (\ref{eq_tautau2}).

5. Evaluate the IB forcing terms at all Lagrangian points using Eqs. (\ref{eq_fib}), (\ref{eq_wib}), (\ref{eq_qib}), and (\ref{eq_fibtang4}).

6. Spread the IB forcing terms to the surrounding Eulerian points using Eq. (\ref{eq_spr2}).

7. Solve Eqs. (\ref{eq_LB}) and (\ref{eq_energy}) including the IB forcing terms and update $\rho$, $\boldsymbol{u}$, $T$, and $p$.

\section{Results and discussions}
\label{res}

In this section, high-Reynolds-number turbulent channel flow is first considered to evaluate the performance of the proposed method for WMLES. Both wall-aligned and wall-unaligned grid configurations are examined. The method is subsequently further validated through simulations of the NACA 23012 airfoil.

\subsection{Turbulent channel flow}
\label{channel}

\begin{figure}
\centerline{\includegraphics[width=1.0\linewidth]{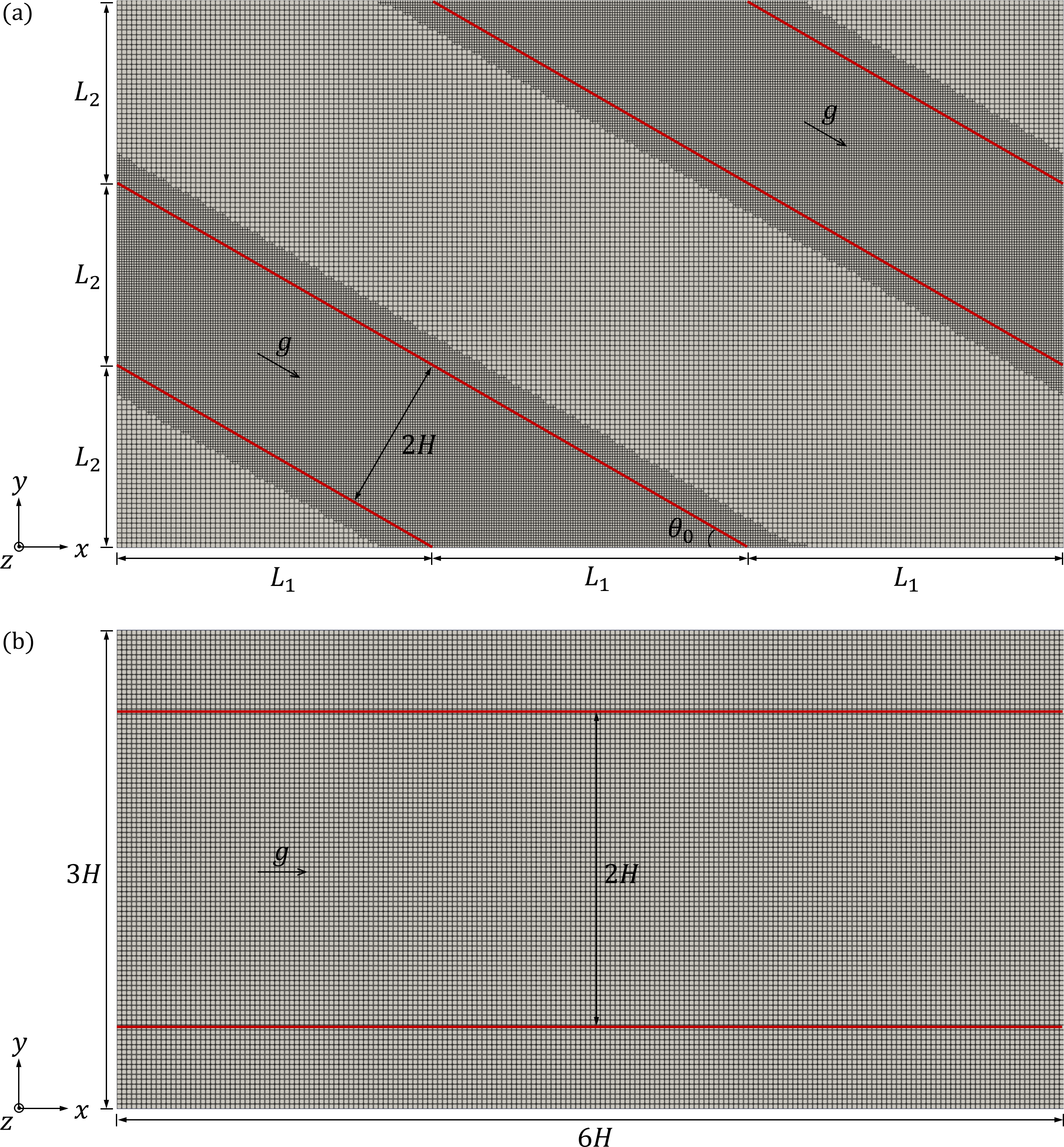}}
\caption{Visualization of the computational domain for turbulent channel flow using (a) a wall-unaligned grid and (b) a wall-aligned grid.}
\label{fig_channel_sche}
\end{figure}

The configuration of the inclined channel with a wall-unaligned grid is illustrated in Fig. \ref{fig_channel_sche}(a). The red line denotes the wall, which is uniformly discretized. The channel height is $2H$, and the inclination angle is $\theta_0$. The computational domain has dimensions of $3L_1 \times 3L_2 \times 4H$, where $L_1$ and $L_2$ are defined as $2H/\sin(\theta_0)$ and $2H/\cos(\theta_0)$, respectively. Periodic boundary conditions are applied on all six boundaries. The fluid motion is resolved on a Cartesian grid that is refined inside the channel and in the near-wall region outside the channel. The ratio between the finest Eulerian grid spacing and the Lagrangian marker spacing is set to $\Delta x/\Delta s = 1$. The flow within the channel is driven by a constant body force $g$. The Reynolds number based on the shear velocity ($u_{\tau}=\sqrt{gH}$) is defined as $Re_{\tau}=u_{\tau}H/\nu$. The Reynolds number ($Re_b=u_b2H/\nu$) based on the bulk velocity ($u_b$) can be estimated by using Dean correlations \cite{dean1978reynolds}:

\begin{equation}
Re_b=\left(\frac{8}{0.073}\right)^{4/7}Re_\tau^{8/7}.
\label{eq_reb}
\end{equation}

A special case of a horizontal channel with a wall-aligned grid is illustrated in Fig. \ref{fig_channel_sche}(b), corresponding to $\theta_0 = 0^{\circ}$. Two walls are located at $y=0$ and $y=2H$, respectively. The computational domain has dimensions of $6H \times 3H \times 4H$ and is discretized using uniform grids.

The flow field is initialized with a uniform flow of velocity $u_b$. After the turbulent flow becomes fully developed, flow statistics are collected over a time interval of $Tu_b/H = 80$ to ensure statistical convergence.

\subsubsection{Assessment of different wall-modeling strategies in WMLES}
\label{sec_difwm}

\begin{table}[t]
\centering
\small
\begin{tabular}{l c c}
  \hline
   Case & Wall-parallel IB forcing & Shear-stress balance \\
  \hline
  No-slip & No-slip condition & \raisebox{-0.2ex}{\Large$\times$} \\
  WS & Wall shear stress & \raisebox{-0.2ex}{\Large$\times$} \\
  WS$+\mu$-model & Wall shear stress & $\mu$-model \\
  WS$+\tau$-model & Wall shear stress & $\tau$-model \\
  \hline
\end{tabular}
\caption{Overview of cases with different wall-modeling strategies.}
\label{table_channel_case}
\end{table}

To assess the performance of the proposed method for WMLES, simulations are performed using different wall-modeling strategies, which are summarized in Table \ref{table_channel_case}. For all cases, an isothermal condition is applied and a non-penetration condition is enforced in the wall-normal direction.

The case WS denotes that the wall-parallel IB force $\boldsymbol{F}^{\mathrm{IB}}_{\xi l}$ is coupled with the wall model through Eq. (\ref{eq_fibtang4}), while the shear-stress balance is not enforced. The case WS+$\mu$-model denotes that $\boldsymbol{F}^{\mathrm{IB}}_{\xi l}$ is coupled with the wall model and the shear-stress balance is enforced using the $\mu$-model defined in Eq. (\ref{eq_taumu}). The case WS+$\tau$-model denotes that $\boldsymbol{F}^{\mathrm{IB}}_{\xi l}$ is coupled with the wall model and the shear-stress balance is enforced using the $\tau$-model defined in Eq. (\ref{eq_tautau2}). A reference case, denoted as No-slip, without any wall-modeling strategy is also included for comparison.

The Reynolds number is fixed at $Re_{\tau}=5200$ ($Re_b \approx 2.6 \times 10^5$), and the inclination angle is set to $\theta_0 = 30^{\circ}$. The reference height and grid spacing are $d_{\mathrm{ref}}=3\Delta x$ and $\Delta x=H/32$, respectively. Consequently, the reference point lies within the inner layer ($d_{\mathrm{ref}} < 0.1\delta_{99}$).

\begin{figure}
\centerline{\includegraphics[width=1.0\linewidth]{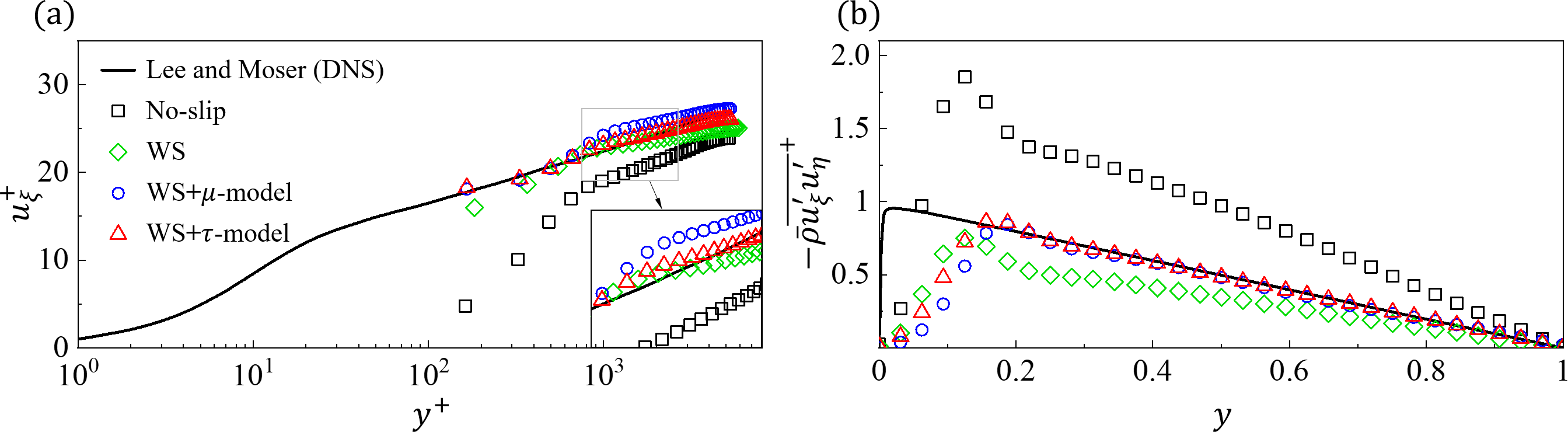}}
\caption{Comparison of different wall-modeling strategies. (a) Mean wall-parallel velocity in wall-viscous units and (b) Reynolds shear stress (FODIBM, $Re_{\tau}=5200$, $\theta_0 = 30 ^{\circ}$, $d_{\mathrm{ref}}=3\Delta x$, $\Delta x=H/32$).}
\label{fig_channel_IBM}
\end{figure}

\begin{table}[t]
\centering
\small
\begin{tabular}{l c c}
  \hline
    & $C_f$ ($\times 10^{-3}$) & $\Delta\epsilon~(\%)$ \\
  \hline
  Lee and Moser (DNS) & 3.442 & - \\
  Present WS & 3.653 & 6.13 \\
  Present WS$+\mu$-model & 3.152 & 8.42 \\
  Present WS$+\tau$-model & 3.422 & 0.58 \\
  \hline
\end{tabular}
\caption{Skin-friction coefficients for WS, WS$+\mu$-model, and WS$+\tau$-model cases (FODIBM, $Re_{\tau}=5200$, $\theta_0 = 30 ^{\circ}$, $d_{\mathrm{ref}}=3\Delta x$, $\Delta x=H/32$).}
\label{table_channel_IBM}
\end{table}

Fig. \ref{fig_channel_IBM} presents the profiles of the mean wall-parallel velocity in wall-viscous units ($\bar{u}^+_{\xi}=\bar{u}_{\xi}/\bar{u}_{\tau}$) and the Reynolds shear stress ($-\bar{\rho}\overline{u_{\xi}^{\prime}u_{\eta}^{\prime}}^+=-\bar{\rho}\overline{u_{\xi}^{\prime}u_{\eta}^{\prime}}/\bar{\tau}_{w}$) obtained from the cases No-slip, WS, WS+$\mu$-model, and WS+$\tau$-model. The value of $\bar{u}^+_{\xi}$ at the wall boundary is not shown. The DNS results of Lee and Moser \cite{lee2015direct} are included for comparison.

For the No-slip case, the mean velocity is significantly underestimated, as shown in Fig. \ref{fig_channel_IBM}(a). This behavior is caused by the insufficient grid resolution near the wall, which is unable to resolve the viscous sublayer, resulting in an underestimated velocity gradient and wall shear stress at the wall ($y=0$). Consequently, the flow velocity in the logarithmic layer decreases. Furthermore, coarse non-body-conforming grids exhibit significant conservation errors in the wall-parallel momentum \cite{capizzano2011turbulent,tamaki2017near,tamaki2021wall}. The inviscid wall-parallel momentum flux through the stepped boundary is non-negligible \cite{tamaki2021wall}, which leads to excessive shear stress entering the channel from the boundary and consequently an overestimation of the Reynolds shear stress, as shown in Fig. \ref{fig_channel_IBM}(b).

For the WS case, the wall shear stress predicted by the wall model is imposed at the wall boundary, which significantly improves both the velocity and Reynolds shear stress profiles compared with the No-slip case. The wall boundary effectively behaves as a slip wall because the no-slip condition is no longer enforced in the wall-parallel direction. In the near-wall region ($y<0.12$ or $y^+<650$), the Reynolds shear stress is not well resolved due to the coarse grid resolution, leading to an underestimation of the Reynolds shear stress throughout the boundary layer. In addition, the velocity gradient in the near-wall region slightly increases relative to the DNS result to compensate for the missing shear stress, resulting in deviations in the mean velocity profile. These observations indicate that additional modeled turbulent shear stress is required.

For the WS+$\mu$-model case, the viscous shear stress is increased by introducing the eddy viscosity $\mu_t^{\mathrm{RANS}}$, which improves both the velocity and Reynolds shear stress profiles compared with the WS case, as shown in Figs. \ref{fig_channel_IBM}(a) and (b). However, a shift of the velocity profile, commonly referred to as the log-layer mismatch, appears when $y^+>700$. This issue is commonly observed in hybrid RANS-LES simulations \cite{cai2023immersed} and is attributed to insufficient wall-normal turbulent mixing in the near-wall region.

In contrast, for the WS+$\tau$-model case, the $\tau$-model mitigates the adverse effects on wall-normal turbulence mixing and effectively eliminates the log-layer mismatch. As shown in Figs. \ref{fig_channel_IBM}(a) and (b), both the velocity and Reynolds shear stress profiles obtained using the WS+$\tau$-model agree well with the DNS results.

To quantify the accuracy of the proposed method, the skin-friction coefficients $C_f = \bar{\tau}_w/(0.5\rho_b u_b^2)$ for the different cases are listed in Table \ref{table_channel_IBM}. The bulk density and velocity are computed as $\rho_b = \int_0^{2H}\bar{\rho}dy/(2H)$ and $u_b = \int_0^{2H}\overline{\rho u}dy/(2\rho_b H)$, respectively. Compared with the DNS result of Lee and Moser \cite{lee2015direct}, the relative errors of $C_f$ obtained using the three wall-modeling strategies are all below $10\%$. Among them, the WS$+\tau$-model yields the most accurate prediction, with a relative error of only $0.58\%$.

\subsubsection{Robustness of the WS\texorpdfstring{$+\tau$}{+tau}-model in WMLES}

To further examine the robustness of the WS+$\tau$-model in WMLES, a sensitivity study is conducted with respect to the grid size ($\Delta x$), reference height ($d_{\mathrm{ref}}$), inclination angle ($\theta_0$), and Reynolds number ($Re_{\tau}$).

\begin{figure}
\centerline{\includegraphics[width=1.0\linewidth]{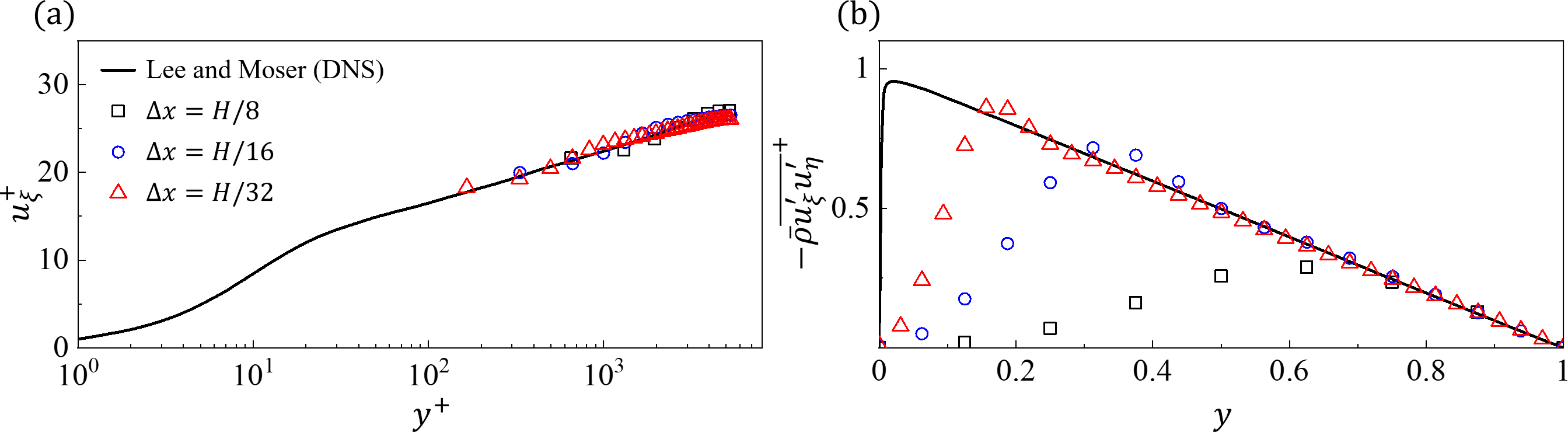}}
\caption{Grid convergence test ($\Delta x$). (a) Mean wall-parallel velocity in wall-viscous units and (b) Reynolds shear stress (FODIBM, WS$+\tau$-model, $Re_{\tau}=5200$, $\theta_0 = 30 ^{\circ}$, $d_{\mathrm{ref}}=3\Delta x$).}
\label{fig_channel_grid}
\end{figure}

\begin{table}[t]
\centering
\small
\begin{tabular}{l c c}
  \hline
    & $C_f$ ($\times 10^{-3}$) & $\Delta\epsilon~(\%)$ \\
  \hline
  Lee and Moser (DNS) & 3.442 & - \\
  Present $\Delta x=H/8$ & 3.332 & 3.19 \\
  Present $H/16$ & 3.493 & 1.48 \\
  Present $H/32$ & 3.422 & 0.58 \\
  \hline
\end{tabular}
\caption{Skin-friction coefficients for different grid resolutions (FODIBM, WS$+\tau$-model, $Re_{\tau}=5200$, $\theta_0 = 30 ^{\circ}$, $d_{\mathrm{ref}}=3\Delta x$).}
\label{table_channel_grid}
\end{table}

First, a grid convergence test is performed. Three grid resolutions are considered: a coarse grid with $\Delta x=H/8$, a medium grid with $\Delta x=H/16$, and a fine grid with $\Delta x=H/32$. Fig. \ref{fig_channel_grid} shows the mean wall-parallel velocity and Reynolds shear stress profiles obtained with different values of $\Delta x$. The velocity profile obtained using the coarse grid exhibits deviations from the DNS results \cite{lee2015direct}. This is mainly because the reference point is located too far from the inner layer ($d_{\mathrm{ref}}=3\Delta x=0.375$), which reduces the accuracy of the wall model. In contrast, the velocity profiles obtained with the medium and fine grids nearly overlap, indicating that spatial convergence has been achieved. Both profiles also agree well with the DNS results. The Reynolds shear stress is not well resolved below the fourth near-wall grid point. However, above this height, it is well predicted even on the coarse grid. As the grid is refined, the Reynolds shear stress shows clear convergence behavior, with finer grids providing more accurate predictions in the near-wall region. The skin-friction coefficients predicted by the present method also agree well with the DNS result, as shown in Table \ref{table_channel_grid}. The maximum relative error is only $3.19\%$ for the coarse grid.

\begin{figure}
\centerline{\includegraphics[width=1.0\linewidth]{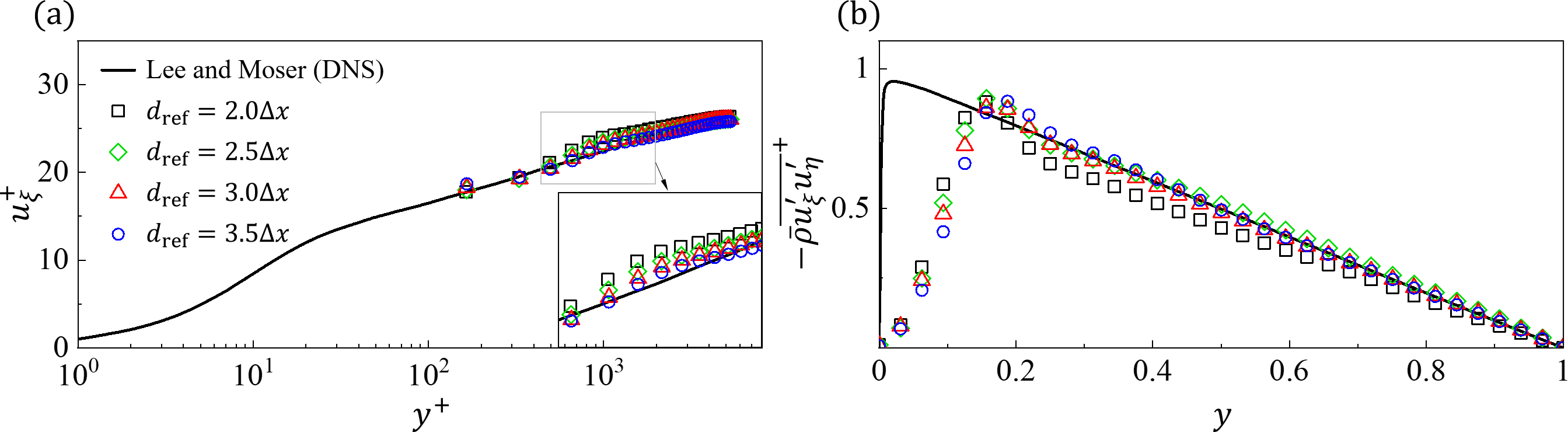}}
\caption{Effect of the reference height ($d_{\mathrm{ref}}$). (a) Mean wall-parallel velocity in wall-viscous units and (b) Reynolds shear stress (FODIBM, WS$+\tau$-model, $Re_{\tau}=5200$, $\theta_0 = 30 ^{\circ}$, $\Delta x=H/32$).}
\label{fig_channel_ref}
\end{figure}

\begin{table}[t]
\centering
\small
\begin{tabular}{l c c}
  \hline
    & $C_f$ ($\times 10^{-3}$) & $\Delta\epsilon~(\%)$ \\
  \hline
  Lee and Moser (DNS) & 3.442 & - \\
  Present $d_{\mathrm{ref}}=2.0\Delta x$ & 3.313 & 3.75 \\
  Present $2.5\Delta x$ & 3.405 & 1.07 \\
  Present $3.0\Delta x$ & 3.422 & 0.58 \\
  Present $3.5\Delta x$ & 3.490 & 1.39 \\
  \hline
\end{tabular}
\caption{Skin-friction coefficients for different reference heights (FODIBM, WS$+\tau$-model, $Re_{\tau}=5200$, $\theta_0 = 30 ^{\circ}$, $\Delta x=H/32$).}
\label{table_channel_ref}
\end{table}

The effect of the reference height ($d_{\mathrm{ref}}$) is further investigated. For $d_{\mathrm{ref}}=2.0\Delta x$, a log-layer mismatch can be observed in Fig. \ref{fig_channel_ref}(a), and the Reynolds shear stress is underestimated above the fourth near-wall grid point as shown in Fig. \ref{fig_channel_ref}(b). Two main reasons account for this behavior. First, when the reference point is too close to the wall, the prediction accuracy of the wall model is inevitably influenced by numerical and subgrid-scale modeling errors \cite{kawai2012wall}. Second, the modeled turbulent shear stress introduced in the near-wall region is insufficient when $d_{\mathrm{ref}}$ is small. As $d_{\mathrm{ref}}$ increases, these issues are gradually mitigated. Correspondingly, the relative error of the skin-friction coefficient also decreases significantly as $d_{\mathrm{ref}}$ increases from $2.0\Delta x$ to $2.5\Delta x$, as shown in Table \ref{table_channel_ref}.

\begin{figure}
\centerline{\includegraphics[width=1.0\linewidth]{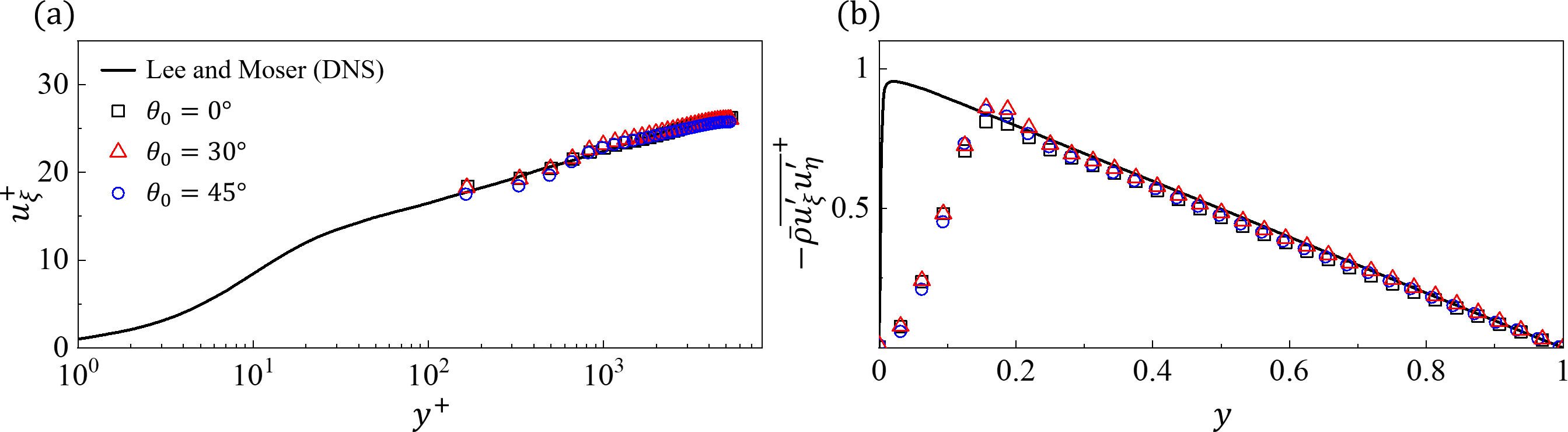}}
\caption{Effect of the inclination angle ($\theta_0$). (a) Mean wall-parallel velocity in wall-viscous units and (b) Reynolds shear stress (FODIBM, WS$+\tau$-model, $Re_{\tau}=5200$, $d_{\mathrm{ref}}=3\Delta x$, $\Delta x=H/32$).}
\label{fig_channel_ang}
\end{figure}

\begin{table}[t]
\centering
\small
\begin{tabular}{l c c}
  \hline
    & $C_f$ ($\times 10^{-3}$) & $\Delta\epsilon~(\%)$ \\
  \hline
  Lee and Moser (DNS) & 3.442 & - \\
  Present $\theta_0=0 ^{\circ}$ & 3.445 & 0.08 \\
  Present $30 ^{\circ}$ & 3.422 & 0.58 \\
  Present $45 ^{\circ}$ & 3.521 & 2.29 \\
  \hline
\end{tabular}
\caption{Skin-friction coefficients for different inclination angles (FODIBM, WS$+\tau$-model, $Re_{\tau}=5200$, $d_{\mathrm{ref}}=3\Delta x$, $\Delta x=H/32$).}
\label{table_channel_ang}
\end{table}

As mentioned earlier, non-body-conforming grids exhibit conservation errors in the wall-parallel momentum across the stepped boundary, which may influence the accuracy of wall modeling. Therefore, it is necessary to examine the effect of the inclination angle ($\theta_0$). As shown in Fig. \ref{fig_channel_ang}, the wall-parallel velocity and Reynolds shear stress profiles obtained for different inclination angles ($0^{\circ}$, $30^{\circ}$, and $45^{\circ}$) almost overlap. Furthermore, the corresponding skin-friction coefficients are listed in Table \ref{table_channel_ang}. The skin-friction coefficient for the wall-aligned grid case ($\theta_0=0^{\circ}$) agrees exactly with the DNS result \cite{lee2015direct}. As $\theta_0$ increases, the relative error ($\Delta\epsilon$) of the skin-friction coefficient increases slightly. The maximum $\Delta\epsilon$, observed at $\theta_0=45^{\circ}$, is only $2.29\%$. These results demonstrate the robustness of the present method for wall modeling over different stepped boundary configurations.

\begin{figure}
\centerline{\includegraphics[width=1.0\linewidth]{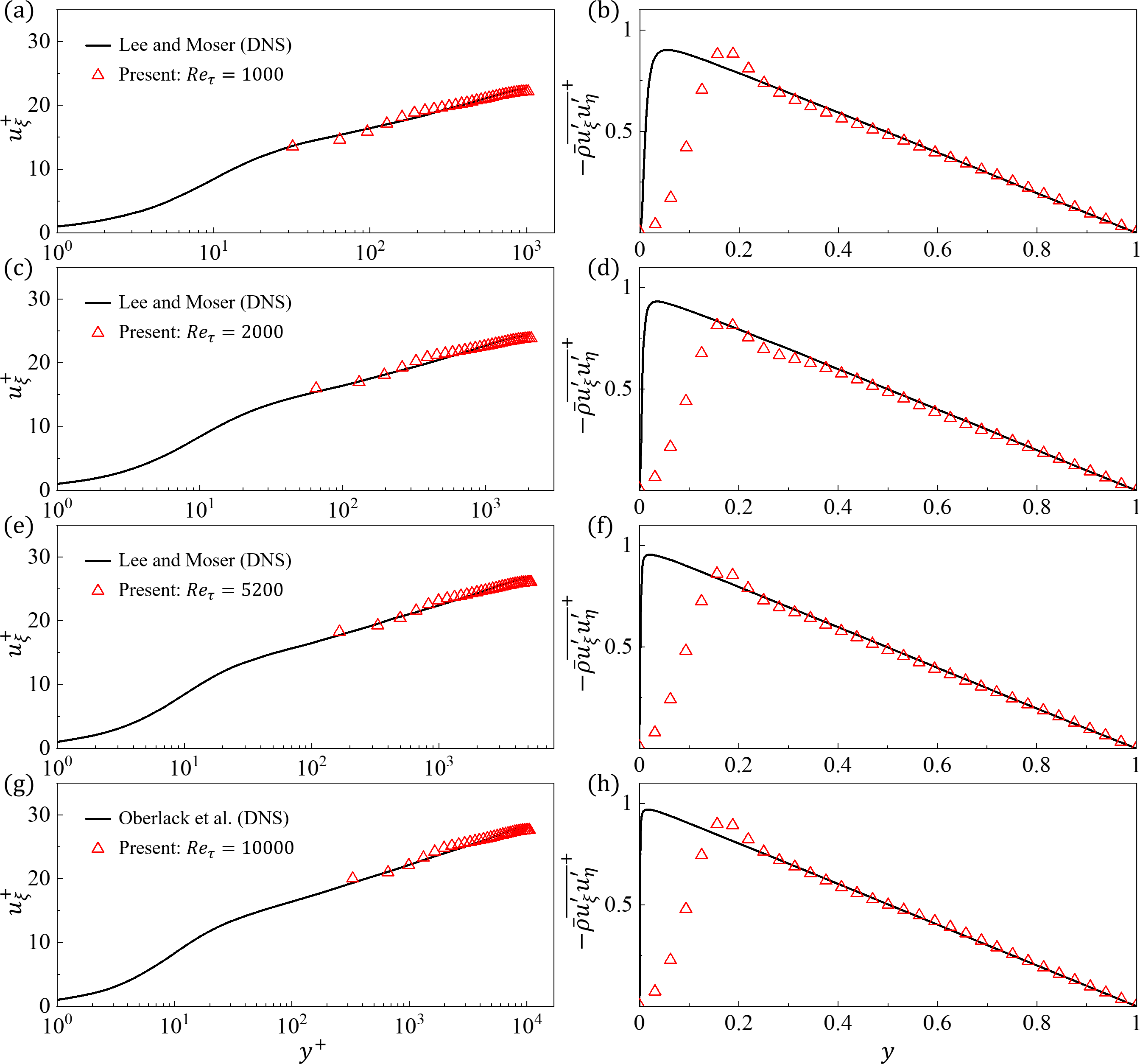}}
\caption{Reynolds number dependence ($Re_\tau$). (a) Mean wall-parallel velocity in wall-viscous units and (b) Reynolds shear stress (FODIBM, WS$+\tau$-model, $\theta_0 = 30 ^{\circ}$, $d_{\mathrm{ref}}=3\Delta x$, $\Delta x=H/32$).}
\label{fig_channel_Re}
\end{figure}

\begin{table}[t]
\centering
\small
\begin{tabular}{l c c}
  \hline
    & $C_f$ ($\times 10^{-3}$) & $\Delta\epsilon~(\%)$ \\
  \hline
  Lee and Moser (DNS) & 5.005 & - \\
  Present $Re_{\tau}=1000$ & 5.001 & 0.08 \\
  Lee and Moser (DNS) & 4.210 & - \\
  Present $Re_{\tau}=2000$ & 4.171 & 0.92 \\
  Lee and Moser (DNS) & 3.442 & - \\
  Present $Re_{\tau}=5200$ & 3.422 & 0.58 \\
  Oberlack et al. (DNS) & 3.050 & - \\
  Present $Re_{\tau}=10000$ & 3.008 & 1.37 \\
  \hline
\end{tabular}
\caption{Skin-friction coefficients for different Reynolds numbers (FODIBM, WS$+\tau$-model, $\theta_0 = 30 ^{\circ}$, $d_{\mathrm{ref}}=3\Delta x$, $\Delta x=H/32$).}
\label{table_channel_Re}
\end{table}

Finally, a series of simulations at different Reynolds numbers ($Re_{\tau}=1000$, 2000, 5200, and 10000) are performed to assess the Reynolds-number dependence of the present method. As shown in Fig. \ref{fig_channel_Re} and Table \ref{table_channel_Re}, the results agree well with the DNS data \cite{lee2015direct,oberlack2021dns} for all considered Reynolds numbers. This indicates that the present WS$+\tau$-model method maintains its accuracy and robustness across a wide range of Reynolds numbers.

\subsubsection{Comparison between DIBM and FODIBM}

To demonstrate the advantages of the FODIBM for wall modeling compared with the conventional DIBM, several simulations using the DIBM are performed. The only difference between the DIBM and FODIBM simulations lies in the interpolation and spreading strategies. Based on the results in Section \ref{sec_difwm}, the WS$+\tau$-model is adopted.

\begin{figure}
\centerline{\includegraphics[width=1.0\linewidth]{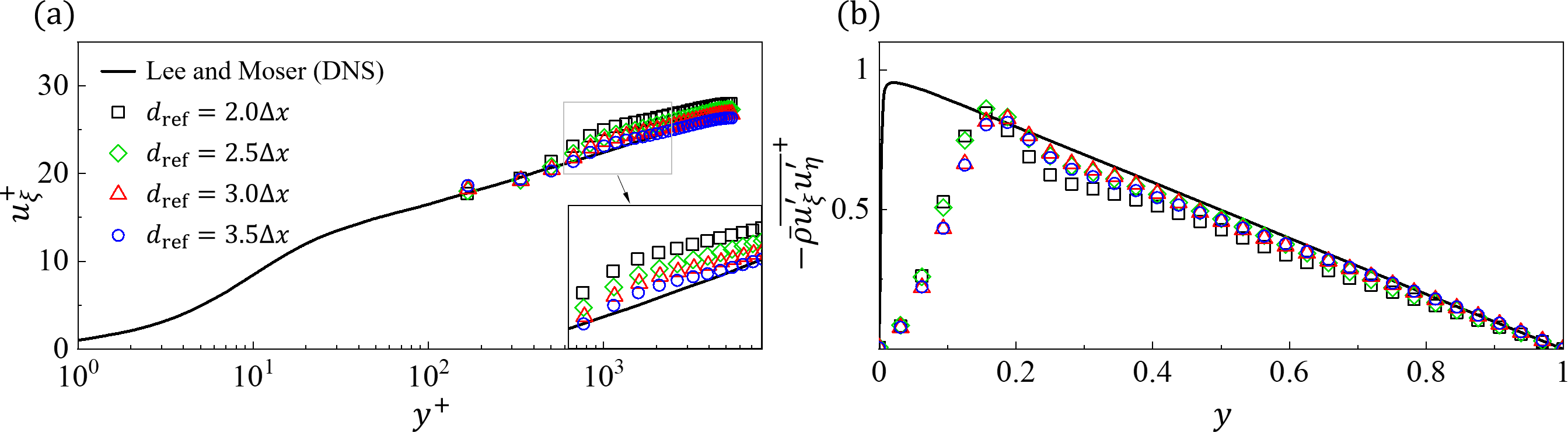}}
\caption{Effect of the reference height ($d_{\mathrm{ref}}$) in DIBM. (a) Mean wall-parallel velocity in wall-viscous units and (b) Reynolds shear stress for different reference height (WS$+\tau$-model, $Re_{\tau}=5200$, $\theta_0 = 30 ^{\circ}$, $\Delta x=H/32$).}
\label{fig_channel_refDIBM}
\end{figure}

\begin{table}[t]
\centering
\small
\begin{tabular}{l c c c c}
  \hline
    & $u_b$ (DIBM) & $\Delta\epsilon~(\%)$ & $u_b$ (FODIBM) & $\Delta\epsilon~(\%)$ \\
  \hline
  Present $d_{\mathrm{ref}}=2.0\Delta x$ & 26.49 & 6.60 & 24.82 & 0.12 \\
  Present $2.5\Delta x$ & 25.93 & 4.35 & 24.62 & 0.92 \\
  Present $3.0\Delta x$ & 25.61 & 3.26 & 24.64 & 0.84 \\
  Present $3.5\Delta x$ & 25.08 & 0.92 & 24.44 & 1.65 \\
  \hline
\end{tabular}
\caption{Bulk velocity ($u_b$) obtained by DIBM and FODIBM with different reference heights (WS$+\tau$-model, $Re_{\tau}=5200$, $\theta_0 = 30 ^{\circ}$, $\Delta x=H/32$). The relative error is computed based on the present results and the theoretical value of 24.85 obtained from an empirical formula.}
\label{table_channel_um}
\end{table}

\begin{table}[t]
\centering
\small
\begin{tabular}{l c c}
  \hline
    & $C_f$ ($\times 10^{-3}$) & $\Delta\epsilon~(\%)$ \\
  \hline
  Lee and Moser (DNS) & 3.442 & - \\
  Present $d_{\mathrm{ref}}=2.0\Delta x$ & 3.004 & 12.7 \\
  Present $2.5\Delta x$ & 3.162 & 8.13 \\
  Present $3.0\Delta x$ & 3.281 & 4.68 \\
  Present $3.5\Delta x$ & 3.382 & 1.74 \\
  \hline
\end{tabular}
\caption{Skin-friction coefficients for different reference heights (DIBM, WS$+\tau$-model, $Re_{\tau}=5200$, $\theta_0 = 30 ^{\circ}$, $\Delta x=H/32$).}
\label{table_channel_refDIBM}
\end{table}

Fig. \ref{fig_channel_refDIBM} shows the wall-parallel velocity and Reynolds shear stress profiles obtained using the DIBM for different values of $d_{\mathrm{ref}}$. For $d_{\mathrm{ref}}=2.0\Delta x$, a noticeable log-layer mismatch is observed and the Reynolds shear stress is underestimated. These discrepancies are more pronounced than those obtained using the FODIBM, as shown in Fig. \ref{fig_channel_ref}. To further quantify the log-layer mismatch in the DIBM and FODIBM results, the bulk velocities predicted by both methods are listed in Table \ref{table_channel_um}. The theoretical value of $24.85$, obtained from $u_b = Re_b \nu /(2H)$, is taken as the reference, where $Re_b$ is estimated using Eq. (\ref{eq_reb}). The relative error of $u_b$ predicted by the DIBM is considerably larger than that obtained using the FODIBM. Furthermore, the relative error of the skin-friction coefficient ($C_f$) predicted by the DIBM reaches $12.7\%$ for $d_{\mathrm{ref}}=2.0\Delta x$, as shown in Table \ref{table_channel_refDIBM}, which is significantly higher than that obtained using the FODIBM ($3.75\%$), as shown in Table \ref{table_channel_ref}.

As $d_{\mathrm{ref}}$ increases, the accuracy of the DIBM results gradually improves. However, the Reynolds shear stress remains underestimated even when $d_{\mathrm{ref}}$ increases to $3.5\Delta x$. The relative errors of $u_b$ and $C_f$ decrease to acceptable values of $0.92\%$ and $1.74\%$, respectively, only when $d_{\mathrm{ref}}$ reaches $3.5\Delta x$. In contrast, the FODIBM enables accurate predictions of the Reynolds shear stress, bulk velocity, and skin-friction coefficient at a relatively low reference height ($d_{\mathrm{ref}}=2.5\Delta x$). This indicates that the conventional DIBM generally requires higher grid resolution than the FODIBM, since the reference point must remain within the valid region of the wall model while being sufficiently separated from the wall.

These observations are attributed to the intrinsic characteristics of the DIBM, in which the IB force is spread in the near-wall region, leading to a diffusion effect. As a result, the IB forcing degrades the accuracy of the wall model at the reference points and disturbs the shear-stress balance in the near-wall region. In addition, the wall-shear-stress condition is not accurately enforced at the wall.

Overall, these results demonstrate that the FODIBM alleviates the limitations of the conventional DIBM and facilitates the application of diffuse-interface IBMs to high-Reynolds-number WMLES.

\subsection{NACA23012 airfoil}

\begin{figure}
\centerline{\includegraphics[width=1.0\linewidth]{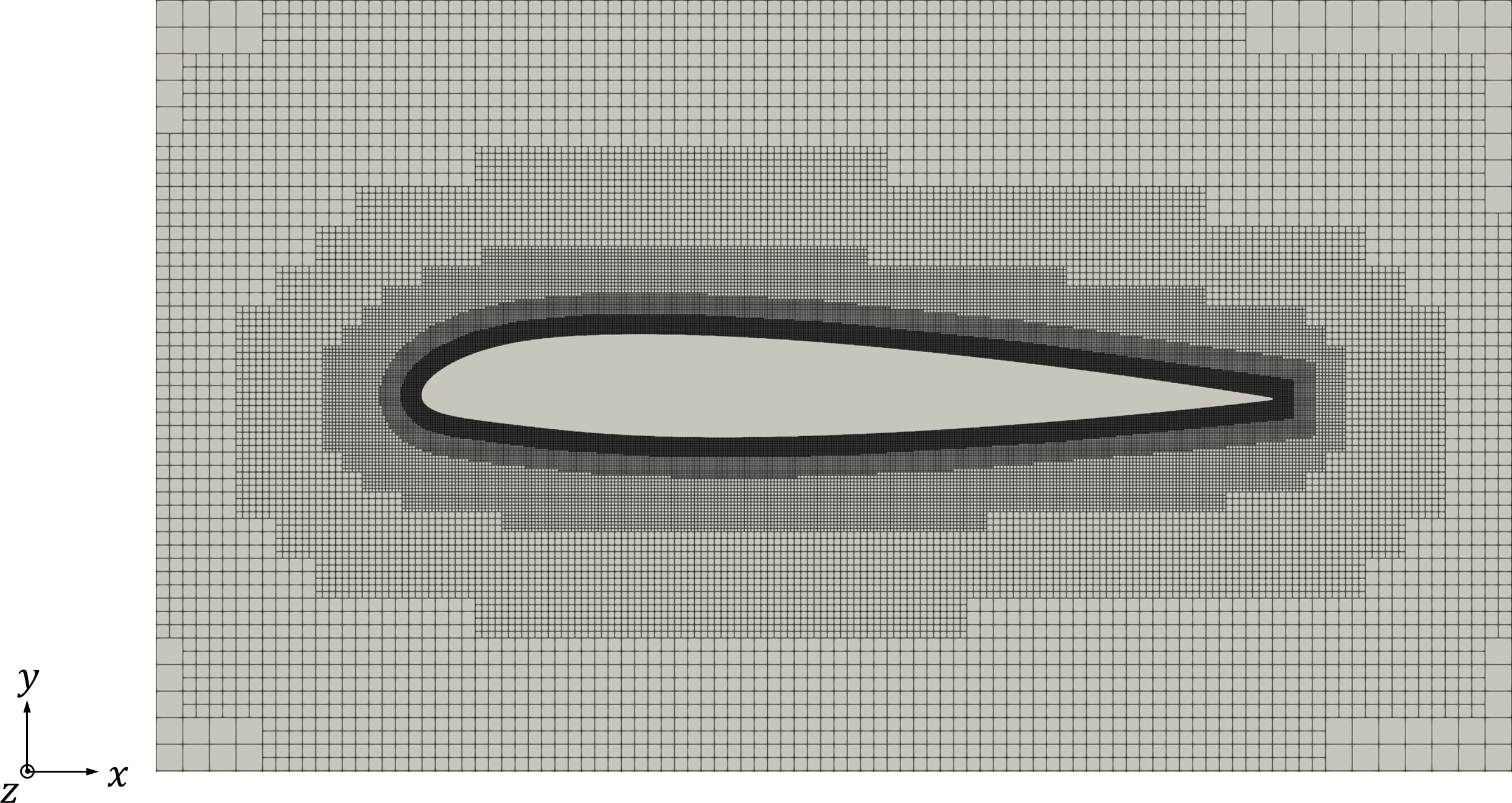}}
\caption{Visualization of the computational mesh in the vicinity of the NACA23012 airfoil.}
\label{fig_naca_sche}
\end{figure}

To further validate the present method in complex configurations relevant to industrial applications, turbulent flow around a NACA23012 airfoil is simulated. The airfoil leading edge is located at $(0,0,0)$ within a rectangular computational domain spanning $[-30c, 60c] \times [-30c, 30c] \times [0, 0.125c]$, where $c$ denotes the chord length. A locally refined mesh is employed, as shown in Fig. \ref{fig_naca_sche}, with minimum grid spacings of $\Delta x_{\mathrm{min}} = c/1024$ and $c/2048$ for the coarse and fine grids, respectively. The averaged value of $y^+$ at the reference points is approximately 220 for $\Delta x_{\mathrm{min}} = c/1024$ and 110 for $\Delta x_{\mathrm{min}} = c/2048$. Periodic boundary conditions are applied in the spanwise ($z$) direction, and absorbing layers are used near the outer boundaries to minimize spurious flow reflections. The free-stream Mach and Reynolds numbers are $Ma_{\infty} = u_{\infty}/c_{\infty} = 0.18$ and $Re = \rho_{\infty} u_{\infty} c / \mu = 1.88 \times 10^6$, respectively, where $u_{\infty}$, $c_{\infty}$, $\rho_{\infty}$, and $\mu$ denote the free-stream velocity, sound speed, density, and dynamic viscosity. The angle of attack is set to $\alpha=6.2^\circ$. The simulation is performed for 30 flow-through times ($c/u_{\infty}$), and statistical averaging is conducted over the last 10 flow-through times to ensure convergence of turbulence statistics.

Due to the under-resolved inner layer in WMLES, it is difficult to obtain fully developed turbulence near the airfoil surface. Therefore, artificial tripping is required to trigger transition. Otherwise, delayed laminar-to-turbulent transition may lead to significant discrepancies in the skin-friction distribution. Here, the tripping strategy proposed by Schlatter and Örlü \cite{schlatter2012turbulent} for DNS of flat-plate turbulent boundary layers is adopted. The key idea is to introduce a weak random volume force in the wall-normal direction, expressed as

\begin{equation}
g(z,t)=A_t[{1-b(t)}h^i(z)+b(t)h^{i+1}(z)],
\end{equation}
where $b(t)=3p^2-2p^3$, $p=t/t_s-i$, and $i=\mathrm{int}(t/t_s)$, with $\mathrm{int}(\cdot)$ denoting the integer part of the argument. The force amplitude is set to $A_t=0.25$, normalized by $\rho_{\infty}\Delta x/\Delta t^2$. The functions $h^i(z)$ are random harmonic signals with unit amplitude for all wavenumbers below $2\pi/z_s$ and zero amplitude for higher wavenumbers. The forcing function $g(z,t)$ is fully characterized by two parameters: the spanwise scale $z_s$ and the temporal scale $t_s$. In the study of Schlatter and Örlü \cite{schlatter2012turbulent}, $z_s=1.7\delta_0^*$ and $t_s=4\delta_0^*/u_{\infty}$ were used, where $\delta_0^*$ denotes the displacement thickness of the incoming laminar Blasius boundary layer. Since $\delta_0^*$ is extremely small relative to the grid resolution used in WMLES, it is set to $2.4\Delta x$ in the present study to obtain moderate values of $z_s$ and $t_s$. Furthermore, in the study of Schlatter and Örlü \cite{schlatter2012turbulent}, $g(z,t)$ was applied as a Gaussian blob centered around the streamwise position $x_0$ near the wall using an exponential spatial distribution. Such a formulation is not convenient to implement in the present configuration. In the present study, tripping forcing is imposed at $x_0/c=0.05$ on the upper surface and $x_0/c=0.2$ on the lower surface of the airfoil. For each Lagrangian point adjacent to these locations, a spreading point is defined at a distance $\Delta x$ along the interface normal. The forcing $\boldsymbol{g}(z,t)=g(z,t)\boldsymbol{\eta}$ is then evaluated at each spreading point and distributed to the surrounding Eulerian points using Eq. (\ref{eq_spr1}). Here, the radius of the delta function is set to $d=1$.

\begin{figure}
\centerline{\includegraphics[width=1.0\linewidth]{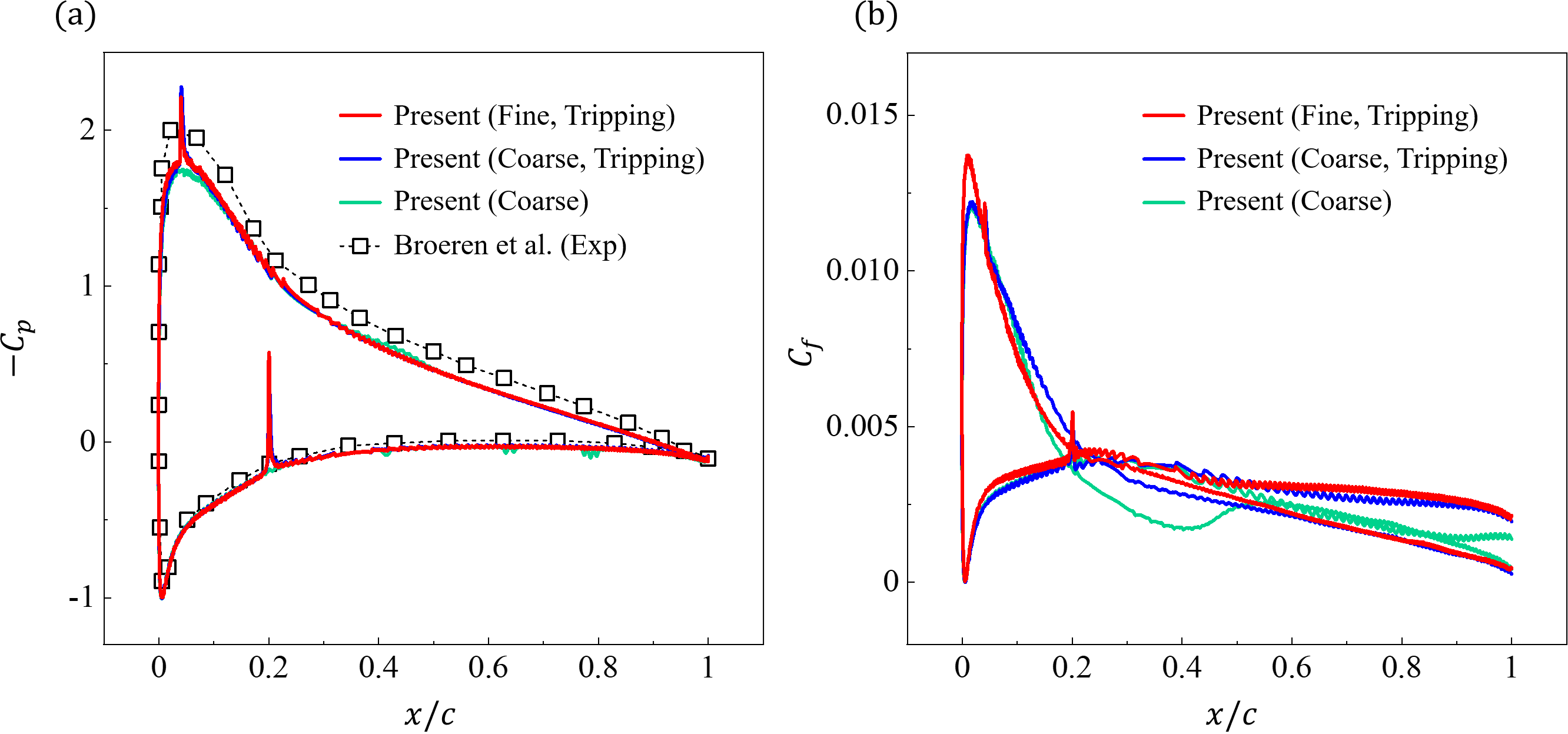}}
\caption{Distributions of (a) pressure coefficient $C_p$ and (b) skin-friction coefficient $C_f$ on the NACA23012 airfoil surface.}
\label{fig_naca_cpcf}
\end{figure}

\begin{figure}
\centerline{\includegraphics[width=1.0\linewidth]{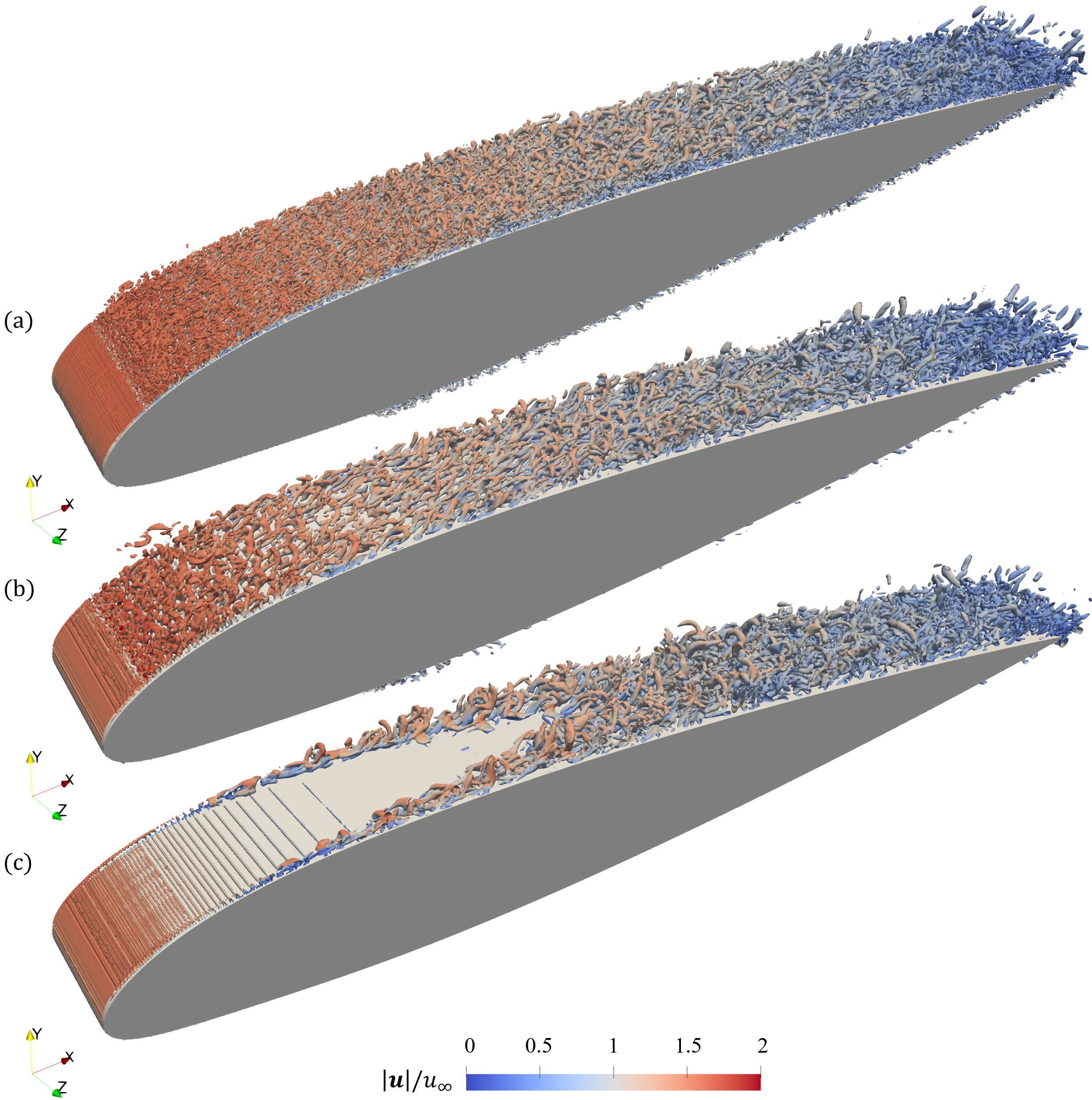}}
\caption{Isosurfaces of the Q-criterion ($Q c^2/u_{\infty}^2 = 1000$) colored by velocity magnitude for (a) the fine grid with tripping, (b) the coarse grid with tripping, and (c) the coarse grid without tripping.}
\label{fig_naca_q}
\end{figure}

Figure \ref{fig_naca_cpcf}(a) compares the distributions of the pressure coefficient $C_p=(\bar{p}-p_{\infty})/(0.5\rho_{\infty}u_{\infty}^2)$ on the airfoil surface between the present results and the reference experimental data \cite{broeren2014validation}. Overall, the present results obtained on both the coarse and fine grids agree well with the reference data. The slightly underestimated $C_p$ on the upper surface may be related to the treatment of the thin laminar boundary layer near the leading edge, where the wall model may be less suitable. This issue is not investigated in detail here, as it is beyond the scope of the present study. The treatment of laminar regions in airfoil flows within the WMLES framework remains an active topic of ongoing research \cite{hussain2025leading}. The peaks observed at $x/c=0.05$ and $0.2$ are caused by the tripping force and cannot be completely avoided. The tripping force only reduces the local accuracy of $C_p$ near these locations, as indicated by the comparison between results obtained with and without tripping. In contrast, Fig. \ref{fig_naca_cpcf}(b) shows that the tripping force significantly influences the prediction of the skin-friction coefficient $C_f=\bar{\tau}_w/(0.5\rho_{\infty}u_{\infty}^2)$. In particular, without tripping, $C_f$ is underestimated in the region $0.2 < x/c < 0.5$ on the upper surface and in the region $0.5 < x/c < 1$ on the lower surface. This behavior is closely related to the development of turbulence near the airfoil surface.

Figure \ref{fig_naca_q} presents the isosurfaces of the Q-criterion colored by velocity magnitude. For the coarse-grid case without tripping, the laminar-to-turbulent transition is delayed. This issue is significantly alleviated after introducing the tripping force. Even on the coarse grid, the tripping strategy enables a more realistic transition behavior. The present method clearly resolves the associated vortical structures during transition. This explains the improved prediction of the skin-friction distribution shown in Fig. \ref{fig_naca_cpcf}(b). The present tripping strategy is more effective in triggering transition than prescribing turbulent fluctuations at the inlet. In addition, inlet turbulence forcing may adversely affect the global accuracy of the $C_p$ prediction.

To further evaluate the aerodynamic-force prediction, the lift coefficient $C_l$ is summarized in Table \ref{table_naca}. In the present study, $C_l=(-\bar{F}^{\mathrm{IB}}_x \mathrm{sin}\alpha+\bar{F}^{\mathrm{IB}}_y \mathrm{cos}\alpha)/(0.5\rho_{\infty}u_{\infty}^2)$ is directly computed from the IB force, where $\boldsymbol{F}^{\mathrm{IB}}=\boldsymbol{F}^{\mathrm{IB}}_{\xi}+\boldsymbol{F}^{\mathrm{IB}}_{\eta}$. The present results agree well with the reference data, with relative errors below $5\%$ for both the fine- and coarse-grid cases. In particular, $C_l$ obtained on the coarse grid with and without tripping are similar, confirming that the effect of the tripping force on the global aerodynamic force is negligible.

\begin{table}[t]
\centering
\small
\begin{tabular}{l c c}
  \hline
    & $C_l$ & $\Delta\epsilon~(\%)$ \\
  \hline
  Present (Fine, Tripping) & 0.760 & 3.7 \\
  Present (Coarse, Tripping) & 0.756 & 4.3 \\
  Present (Coarse) & 0.754 & 4.5 \\
  Broeren et al. (Exp) & 0.790 & - \\
  \hline
\end{tabular}
\caption{Comparison of lift ($C_l$) coefficient.}
\label{table_naca}
\end{table}

\section{Conclusions}
\label{conclu}

This work proposes an approach for wall-modeled large eddy simulation (WMLES) based on a fully one-sided diffuse-interface immersed boundary method (FODIBM) \cite{mao2026explicit}. In the FODIBM, both interpolation and spreading are performed exclusively inside the immersed boundary. First, an approach (WS) is proposed to enforce the wall-shear-stress condition by coupling the wall-parallel IB forcing with the wall shear stress predicted by an explicit wall model, without introducing any artificial parameters. The wall-normal IB forcing, defined in the conventional manner, is used to enforce the non-penetration condition in the wall-normal direction. Second, two approaches are considered to ensure the shear-stress balance below the reference height, namely the $\mu$-model based on the modeled eddy viscosity and the $\tau$-model based on the modeled turbulent shear-stress tensor.

The proposed method is first validated using high-Reynolds-number turbulent channel flows. Among the tested strategies, the WS$+\tau$-model provides the most accurate predictions of the mean velocity and Reynolds shear stress profiles, as well as the skin-friction coefficient, all of which show good agreement with the DNS results. By imposing an appropriate wall shear stress at the boundary through the WS treatment, the significant discrepancies in the mean velocity and Reynolds shear stress profiles observed in the no-slip case are greatly reduced. However, slight deviations remain when only WS is applied. Enforcing shear-stress balance through the $\mu$-model or $\tau$-model is therefore essential to further correct the logarithmic velocity profile and compensate for the under-resolved Reynolds shear stress. Compared with the $\mu$-model, the $\tau$-model mitigates adverse effects on wall-normal turbulence mixing and effectively eliminates the log-layer mismatch. The robustness of the proposed WS$+\tau$-model is further assessed by examining the effects of grid size, reference height, inclination angle, and Reynolds number. The results demonstrate that the proposed WS$+\tau$-model maintains good accuracy and robustness under various conditions. Compared with the conventional diffuse-interface immersed boundary method, the FODIBM enables more accurate predictions, particularly at relatively low reference heights. This improvement is attributed to the reduced diffusion effect of the FODIBM, which otherwise affects the accuracy of the wall model and disturbs the shear-stress balance in the near-wall region. In addition, the performance of the proposed method is further verified through simulations of turbulent flow over a NACA23012 airfoil. The present predictions show
good agreement with the experimental data, further confirming the capability of the proposed method for high-Reynolds-number WMLES.

Although the present method is validated within a lattice Boltzmann method (LBM) solver, it can be straightforwardly extended to Navier-Stokes-based solvers. Future work will focus on more complex industrial applications involving flow separation.

\vspace{2ex} \noindent \textbf{Declaration of competing interest}

\vspace{2ex} The authors declare that they have no known competing financial interests or personal relationships that could have appeared to influence the work reported in this paper.

\vspace{2ex} \noindent \textbf{Acknowledgements}

\vspace{2ex} This research was supported by the FALCON project funded by the European Union's Horizon Europe research and innovation programme under grant agreement No 101138305, and by the ANR, Airbus, Fives-Pillard and SafranTech through the Industrial Chair Program LIBERTY ANR-23-CHIN-0005. Centre de Calcul Intensif d'Aix-Marseille and GENCI-TGCC/CINES (Grant A0152B11951) are acknowledged for granting access to their high-performance computing resources.

\vspace{2ex} \noindent \textbf{Data availability}

\vspace{2ex} Data will be made available on request.

\appendix
\section{D3Q19 Basis}
\label{app_d3q19}

The discrete velocities $\boldsymbol{c}_i$ of the $D3Q19$ scheme are defined as
\begin{equation}
\begin{gathered}
  c_{i,x}=\{0,1,1,0,-1,0,0,0,1,-1,-1,-1,0,1,0,0,0,-1,1\},\\
  c_{i,y}=\{0,0,0,0,0,1,1,-1,1,1,0,0,0,0,-1,-1,1,-1,-1\},\\
  c_{i,z}=\{0,0,1,1,1,0,1,1,0,0,0,-1,-1,-1,0,-1,-1,0,0\}.
\end{gathered}
\label{eq_cid3q19}
\end{equation}

The lattice weights $\omega_{i}$ of the $D3Q19$ scheme are given by
\begin{equation}
\begin{split}
  \omega_{i}=\left\{\frac{1}{3},\frac{1}{18},\frac{1}{36},\frac{1}{18},\frac{1}{36},\frac{1}{18},\frac{1}{36},\frac{1}{36},\frac{1}{36},\frac{1}{18},\frac{1}{36},\frac{1}{18},\frac{1}{18},\frac{1}{36},\right.\\
  \left.\frac{1}{18},\frac{1}{36},\frac{1}{36},\frac{1}{36},\frac{1}{36}\right\}.
\end{split}
\label{eq_wi}
\end{equation}

The equilibrium distribution function $f^{eq}_i$ is expressed as

\begin{equation}
    \begin{aligned}
        f_i^{eq} = \omega_i \bigg[ \rho 
            &+ \frac{\omega_i - \delta_{0i}}{\omega_i} \rho (\theta - 1) + \frac{\mathcal{H}_{i,\alpha}^{(1)}}{c_s^2} \rho u_\alpha + \frac{\mathcal{H}_{i,\alpha\beta}^{(2)}}{2c_s^4} 
            \rho u_\alpha u_\beta \\
            &+ \frac{\mathcal{H}_{i,\alpha\beta\gamma}^{(3r)}}{6c_s^6} 
            \rho u_\alpha u_\beta u_\gamma \bigg],
    \end{aligned}
\label{eq_feq}
\end{equation}
where $\theta= T/T_{ref}$ is the normalized temperature obtained from the resolution of the energy equation (\ref{eq_energy}). $T_{ref}$ is an arbitrary reference temperature used to adjust the CFL number \cite{wissocq2022restoring}. $c_s=\sqrt{RT_{ref}}=\Delta x/(\sqrt{3}\Delta t)$ is the lattice sound speed. $\mathcal{H}_i$ are discrete Hermite polynomials defined in Eqs. (\ref{eq_hi}) and (\ref{eq_hi3}).

The off-equilibrium distribution function $\bar{f}^{neq}_i$ is calculated by

\begin{equation}
\bar{f}_i^{neq}=\omega_i\left\{\frac{\mathcal{H}_{i,\alpha\beta}^{(2)}}{2c_s^4}\Pi_{\alpha\beta}^{{neq},(2)}+\frac{\mathcal{H}_{i,\alpha\beta\gamma}^{(3r)}}{6c_s^6}\Pi_{\alpha\beta\gamma}^{{neq},(3r)}\right\},
\label{eq_fneq}
\end{equation}
where $\Pi_{\alpha\beta}^{{neq},(2)}$ and $\Pi_{\alpha\beta\gamma}^{{neq},(3)}$ denote second- and third-order Hermite off-equilibrium coefficients, defined in Eqs. (\ref{eq_pi2}) and (\ref{eq_pi3}). These coefficients are evaluated using the hybrid recursive collision model proposed in \cite{jacob2018new}, which improves numerical stability.

The discrete Hermite polynomials $\mathcal{H}_i$ are calculated by
\begin{equation}
\begin{gathered}
    \mathcal{H}_i^{(0)} = 1,\\
    \mathcal{H}_{i,\alpha}^{(1)} = c_{i\alpha},\\
    \mathcal{H}_{i,\alpha\beta}^{(2)} = c_{i\alpha}c_{i\beta} - c_s^2\delta_{\alpha\beta},\\
    \mathcal{H}_{i,\alpha\beta\gamma}^{(3)} = c_{i\alpha}c_{i\beta}c_{i\gamma} - c_s^2(\delta_{\alpha\beta}c_{i\gamma} + \delta_{\beta\gamma}c_{i\alpha} + \delta_{\alpha\gamma}c_{i\beta}),
\end{gathered}
\label{eq_hi}
\end{equation}

\begin{equation}
\begin{gathered}
    \mathcal{H}_{i,1}^{(3r)} = \mathcal{H}_{i,xxy}^{(3)} + \mathcal{H}_{i,yzz}^{(3)},\\
    \mathcal{H}_{i,2}^{(3r)} = \mathcal{H}_{i,xzz}^{(3)} + \mathcal{H}_{i,xyy}^{(3)},\\
    \mathcal{H}_{i,3}^{(3r)} = \mathcal{H}_{i,yyz}^{(3)} + \mathcal{H}_{i,xxz}^{(3)},\\
    \mathcal{H}_{i,4}^{(3r)} = \mathcal{H}_{i,xxy}^{(3)} - \mathcal{H}_{i,yyz}^{(3)},\\
    \mathcal{H}_{i,5}^{(3r)} = \mathcal{H}_{i,xzz}^{(3)} - \mathcal{H}_{i,xyy}^{(3)},\\
    \mathcal{H}_{i,6}^{(3r)} = \mathcal{H}_{i,yyz}^{(3)} - \mathcal{H}_{i,xxz}^{(3)}.
\end{gathered}
\label{eq_hi3}
\end{equation}

The second-order Hermite off-equilibrium coefficient $ \mathrm{\Pi}_{\alpha\beta}^{{neq},(2)}$ is defined as
\begin{equation}
    \begin{aligned}
        \mathrm{\Pi}_{\alpha\beta}^{{neq},(2)}(x,t) = 
        &\ \tau \sum_{i=0}^{Q-1} 
        \left[
            c_{i\alpha} c_{i\beta} - \frac{\delta_{\alpha\beta}}{3} c_{i\gamma} c_{i\gamma}
        \right] \\
        &\left( 
            \bar{f}_i(x,t) - f^{eq}(x,t) + \frac{\Delta t}{2} F_i(x,t - \Delta t)
        \right) -\\
        & \Bigg[ 
            (1 - \tau)\rho c_s^2 \bar{\tau}
            \left(
                \frac{\partial u_\alpha}{\partial x_\beta} + \frac{\partial u_\beta}{\partial x_\alpha}
                - \frac{2\delta_{\alpha\beta}}{D} \frac{\partial u_\gamma}{\partial x_\gamma}
            \right)
        \Bigg],
    \end{aligned}
\label{eq_pi2}
\end{equation}
where $\tau$ is the weighting free parameter \cite{jacob2018new}, which is set to 0.98 in the present study. $D$ is the spatial dimension. The third-order Hermite off-equilibrium coefficient $\Pi_{\alpha\beta\gamma}^{{neq},(3)}$ is then calculated by
\begin{equation}
\Pi_{\alpha\beta\gamma}^{{neq},(3)}(x,t)=\left[u_\alpha\Pi_{\beta\gamma}^{{neq},(2)}+u_\beta\Pi_{\gamma\alpha}^{{neq},(2)}+u_\gamma\Pi_{\alpha\beta}^{{neq},(2)}\right](x,t).
\label{eq_pi3}
\end{equation}

The Hermite moment $a_{\alpha\beta}^{F,(2)}$ is defined as
\begin{equation}
\begin{aligned}
a_{\alpha\beta}^{F,(2)}=&\frac{2}{D}\delta_{\alpha\beta}\rho c_{s}^{2}\frac{\partial u_{\gamma}}{\partial x_{\gamma}}-\delta_{\alpha\beta}c_{s}^{2}\frac{\partial\rho(1-\theta)}{\partial t}+a_{\alpha\beta}^{C}+f_{u,\alpha}u_{\beta}+f_{u,\beta}u_{\alpha},
\end{aligned}
\label{eq_af2}
\end{equation}
where $a_{\alpha\beta}^{C}$ is the lattice-dependent component of the force term introduced in \cite{farag2021unified}.

The discrete operators $\delta_t$ and $\delta_\alpha$ are defined as
\begin{equation}
\begin{gathered}
\delta_t\Phi=\frac{\Phi(\boldsymbol{x},t+\Delta t)-\Phi(\boldsymbol{x},t)}{\Delta t},\\
\delta_\alpha\Phi=\frac{\Phi(\boldsymbol{x},t)-\Phi(\boldsymbol{x}-\boldsymbol{e}_\alpha\Delta x,t)}{\Delta x},
\end{gathered}
\label{eq_deltata}
\end{equation}
where $\boldsymbol{e}_\alpha$ is the unity vector in the direction $\alpha$.

The mass flux $F_{+\Delta \alpha/2}^{\rho}$ are defined by
\begin{equation}
\begin{aligned}
&F_{+\Delta x/2}^{\rho}(x,y,z)=\frac{\Delta x}{\Delta t}\Big[f_{1}^{col}(x,y,z)-f_{10}^{col}(x^{+},y,z)\\&+\frac12f_2^{col}(x,y,z^-)-\frac12f_{11}^{col}(x^+,y,z)+\frac12f_2^{col}(x,y,z)\\&-\frac12f_{11}^{col}(x^+,y,z^+)-\frac12f_4^{col}(x^+,y,z^-)+\frac12f_{13}^{col}(x,y,z)\\&-\frac12f_4^{col}(x^+,y,z)+\frac12f_{13}^{col}(x,y,z^+)+\frac12f_8^{col}(x,y^-,z)\\&-\frac12f_{17}^{col}(x^+,y,z)+\frac12f_8^{col}(x,y,z)-\frac12f_{17}^{col}(x^+,y^+,z)\\&-\frac12f_9^{col}(x^+,y^-,z)+\frac12f_{18}^{col}(x,y,z)-\frac12f_9^{col}(x^+,y,z)\\&+\frac12f_{18}^{col}(x,y^+,z)\Big],
\end{aligned}
\label{eq_fluxx}
\end{equation}

\begin{equation}
\begin{aligned}
&F_{+\Delta y/2}^{\rho}(x,y,z)=\frac{\Delta x}{\Delta t}\Big[f_{5}^{col}(x,y,z)-f_{14}^{col}(x,y^{+},z)\\&+\frac12f_8^{col}(x,y,z)-\frac12f_{17}^{col}(x^+,y^+,z)+\frac12f_8^{col}(x^-,y,z)\\&-\frac12f_{17}^{col}(x,y^+,z)+\frac12f_9^{col}(x,y,z)-\frac12f_{18}^{col}(x^-,y^+,z)\\&+\frac12f_9^{col}(x^+,y,z)-\frac12f_{18}^{col}(x,y^+,z)+\frac12f_6^{col}(x,y,z^-)\\&-\frac12f_{15}^{col}(x,y^+,z)+\frac12f_6^{col}(x,y,z)-\frac12f_{15}^{col}(x,y^+,z^+)\\&-\frac12f_7^{col}(x,y^+,z^-)+\frac12f_{16}^{col}(x,y,z)-\frac12f_7^{col}(x,y^+,z)\\&+\frac12f_{16}^{col}(x,y,z^+)\Big],
\end{aligned}
\label{eq_fluxy}
\end{equation}

\begin{equation}
\begin{aligned}
&F_{+\Delta z/2}^{\rho}(x,y,z)=\frac{\Delta x}{\Delta t}\biggl[f_{3}^{col}(x,y,z)-f_{12}^{col}(x,y,z^{+})\\&+\frac12f_2^{col}(x,y,z)-\frac12f_{11}^{col}(x^+,y,z^+)+\frac12f_2^{col}(x^-,y,z)\\&-\frac12f_{11}^{col}(x,y,z^+)+\frac12f_4^{col}(x,y,z)-\frac12f_{13}^{col}(x^-,y,z^+)\\&+\frac12f_4^{col}(x^+,y,z)-\frac12f_{13}^{col}(x,y,z^+)+\frac12f_6^{col}(x,y,z)\\&-\frac12f_{15}^{col}(x,y^+,z^+)+\frac12f_{6}^{col}(x,y^-,z)-\frac12f_{15}^{col}(x,y,z^+)\\&+\frac12f_7^{col}(x,y,z)-\frac12f_{16}^{col}(x,y^-,z^+)+\frac12f_7^{col}(x,y^+,z)\\&-\frac12f_{16}^{col}(x,y,z^+)\Big],
\end{aligned}
\label{eq_fluxz}
\end{equation}
where $x^\pm=x\pm\Delta x $ and $y^\pm=y\pm\Delta x$. The expressions of the momentum flux $F_{+\Delta \alpha/2}^{\rho u_{\alpha}}$ can then be obtained straightforwardly by replacing each $f^{col}_i$ in $F_{+\Delta \alpha/2}^{\rho}$ by $c_{i,\alpha}f^{col}_i$.

The linear function $\mathscr{F}_{+\Delta\boldsymbol{\alpha}/2}^*$ is given by
\begin{equation}
\mathscr{F}_{+\Delta\boldsymbol{\alpha}/2}^*(\Phi)=\begin{cases}\overline{\Phi}_{+\Delta\boldsymbol{\alpha}/2}(\boldsymbol{x},t) &\mathrm{if}\quad\tilde{\boldsymbol{u}}_{\alpha}\geq0,\\
\overline{\Phi}_{-\Delta\boldsymbol{\alpha}/2}(\boldsymbol{x}+\boldsymbol{e}_{\alpha}\Delta x,t)&\mathrm{else},\end{cases}
\label{eq_linear}
\end{equation}
where $\tilde{\boldsymbol{u}}_{\boldsymbol{\alpha}}$ is defined as $\tilde{\boldsymbol{u}}_{\alpha}=[\boldsymbol{u}_{\alpha}(\boldsymbol{x},t)+\boldsymbol{u}_{\alpha}(\boldsymbol{x}+\boldsymbol{e}_{\alpha}\Delta x,t)]/2$ to ensure the symmetry of the algorithm,
\begin{equation}
\begin{gathered}
\overline{\Phi}_{+\Delta\alpha/2}=\Phi_{+\Delta\alpha/2}+\frac{\boldsymbol{u}_\alpha}{2}\frac{\Delta t}{\Delta x}\left(\Phi_{-\Delta\alpha/2}-\Phi_{+\Delta\alpha/2}\right), \\
\overline{\Phi}_{-\Delta\alpha/2}=\Phi_{-\Delta\alpha/2}+\frac{\boldsymbol{u}_\alpha}{2}\frac{\Delta t}{\Delta x}\left(\Phi_{-\Delta\alpha/2}-\Phi_{+\Delta\alpha/2}\right), \\
\Phi_{+\Delta\alpha/2}=\Phi+\frac{\Delta_{\alpha}}{2},\quad\Phi_{-\Delta\alpha/2}=\Phi-\frac{\Delta_{\alpha}}{2},
\end{gathered}
\end{equation}
where $\Delta_{\alpha}$ approximates the slope of $\Phi$ in the $\alpha$-direction, and is given by
\begin{equation}
\begin{aligned}
\Delta_{\alpha}=&\frac12\Big\{(1+\boldsymbol{\eta}_\alpha)[\Phi(\boldsymbol{x},t)-\Phi(\boldsymbol{x}-\boldsymbol{e}_\alpha\Delta x,t)]\\&+(1-\boldsymbol{\eta}_\alpha)[\Phi(\boldsymbol{x}+\boldsymbol{e}_\alpha\Delta x,t)-\Phi(\boldsymbol{x},t)]\Big\},
\end{aligned}
\end{equation}
where $\boldsymbol{\eta}_{\alpha}$ is defined as $\boldsymbol{\eta}_{\alpha}=\frac{1}{3}\left[2\boldsymbol{u}_{\alpha}\frac{\Delta t}{\Delta x}-\mathrm{sign}(\boldsymbol{u}_{\alpha})\right]$ to obtain  a third-order accurate convection scheme in space and time \cite{wissocq2022restoring}.

\bibliographystyle{elsarticle} 
\bibliography{elsarticle}

\end{document}